\RequirePackage{fix-cm}
\documentclass[twocolumn,epjc3]{svjour3}  
\smartqed  
\RequirePackage{graphicx}
\RequirePackage{mathptmx}      
\def\x{\mathbf{x}}

\newcommand{\e}{\mathrm{e}}

\RequirePackage[colorlinks,citecolor=blue,urlcolor=blue,linkcolor=blue]{hyperref}
\usepackage{amsmath,graphicx,amscd,xcolor}
\usepackage{braket}
\usepackage{verbatim} 
\usepackage{cite}

\journalname{Eur. Phys. J. C}
\begin{document}

\title{Probing thermal fluctuations through scalar test particles}


\author{G. H. S. Camargo\thanksref{e1,addr1}
        \and
        V. A. De Lorenci\thanksref{e2,addr2}
        \and
        A. L. Ferreira Junior\thanksref{e3,addr2,addr5}
        \and
        C. C. H. Ribeiro\thanksref{e3,addr3,addr4} 
}

\thankstext{e1}{e-mail: guilhermehenrique@unifei.edu.br}
\thankstext{e2}{e-mail: delorenci@unifei.edu.br}
\thankstext{e3}{e-mail: alexsandref@unifei.edu.br}
\thankstext{e4}{e-mail: caiocesarribeiro@ifsc.usp.br}


\institute{Instituto de Ci\^encias Exatas, Universidade Federal de Juiz de Fora, Juiz de Fora, Minas Gerais 36036-330,  Brazil \label{addr1}
           \and
           Instituto de F\'{\i}sica e Qu\'{\i}mica, Universidade Federal de Itajub\'a, Itajub\'a, Minas Gerais 37500-903, Brazil \label{addr2}
           \and
           PPGCosmo, CCE - Universidade Federal do Esp\'{\i}rito Santo, Vit\'{o}ria, 29075-910, Brazil.
\label{addr5}
           \and
           Instituto de F\'{\i}sica de S\~ao Carlos, Universidade de S\~ao Paulo, S\~ao Carlos, S\~ao Paulo 15980-900, Brazil \label{addr3}
           \and
           Seoul National University, Department of Physics and Astronomy, Seoul 08826, Korea \label{addr4}
}

\date{Received: date / Accepted: date}

\maketitle

\begin{abstract}
The fundamental vacuum state of quantum fields, related to Minkowski space, produces divergent fluctuations that must be suppressed in order to bring reality to the descrip-tion of physical systems. As a consequence, negative vacuum expectation values of classically positive-defined quantities can appear. This has been addressed in the literature as subvacuum phenomenon. Here it is investigated how a scalar charged test particle is affected by the vacuum fluctuations of a massive scalar field in D+1 spacetime when the background evolves from empty space to a thermal bath, and also when a perfectly reflecting boundary is included.\, It is shown that when the particle is brought into a thermal bath it gains an amount of energy by means of positive dispersions of its velocity components. The magnitude of this effect is dependent on the temperature and also on the field mass. However, when a reflecting wall is inserted, dispersions can be positive or negative, showing that subvacuum effect happens even in a finite temperature environment. Furthermore, a remarkable result is that temperature can even improve negative velocity fluctuations. The magnitude of the residual effects depends on the switching interval of time the system takes to evolve between two states.
\end{abstract}

\section{Introduction}
\label{intro}
The vacuum state of a quantum field is an extremely versatile structure and its fingerprints can be found in a plethora of physical systems, ranging from semiclassical gravity to solid state physics. From the gravitational side, perhaps the most emblematic example is the so-called Hawking radiation \cite{Hawking74}, in which the formation of black holes is predicted to be associated to the emission of thermal radiation. We may also quote the more severe effect of spontaneous scalarization, where the vacuum fluctuations lead to phase transitions in certain spacetimes \cite{Landulfo2015}. 
In the context of a quantum scalar field near a reflecting wall in locally flat spacetime, requirement of thermodynamic equilibrium leads to a restriction on the possible values of the curvature coupling parameter $\xi$  \cite{Delorenci2015,Moreira2017}. Particularly, it is shown that for higher dimensions the minimal coupling is ruled out. 
From the condensed matter perspective, examples include the vacuum polarization near boundaries, like the Casimir effect \cite{Casimir1948}, and the emergence of a superconducting phase in some systems \cite{Cao2018}.

Notice that the above mentioned instances of vacuum-related phenomena are linked to the transition between vacuum states of some field, driven by some external agent. In striking contrast to this, we have the Unruh effect \cite{Unruh1976}, in which a detector can perceive the Minkowski vacuum state as a thermal bath of particles, depending on its state. This remarkable example shows that associated to each manifestation of the quantum vacuum is the problem of how to measure it. Clearly, probing quantum effects is an easy task in some systems, like the emergence of a superconducting phase, where measurements can be made through simple transport properties. However, there exists examples of more elusive phenomena like subvacuum effects, where classically positive-definite observables can assume negative values after renormalization. 

In this regard, fingerprints of quantum field fluctuations were shown to be present on the motion of electric charged particles \cite{ford2004,johnson2002}. There, a charged particle originally still in empty space, gains a stochastic motion when the electromagnetic vacuum is modified by the introduction of  a reflective boundary. Nonetheless, velocity fluctuations seems to diverge at the boundary and when $\tau=2x$, which corresponds to a round trip of a photon between the particle and the boundary. Such divergences are linked back to the punctual nature of the particle, and to idealizations of the boundary, i.e., the perfectly reflectiveness, and its sudden appearance. Since then, this model has been investigated in several possible arrangements \cite{hongwei2006,bessa2009,Seriu2009,seriu2008,delorenci2014,delorenci2016,Camargo2018,delorenci2019,delorenci2019b}. In particular, the divergences naturally led to a systematic study of regularization procedures adapted to this kind of system. For instance, the quantum nature of the particle was studied \cite{Seriu2009}, and results were obtained in some special regimes. Another approach consists in implementing a smooth switching of the interaction, which in some cases can lead to analytic results \cite{delorenci2016,delorenci2019b,Camargo2018,delorenci2019}. It is also noteworthy that negative uncertainties, a fingerprint of subvacuum phenomena, was linked to a decrease in the overall positive quantum uncertainty \cite{ford2004} or the particle kinetic energy \cite{Camargo2018}.  
We also mention a recent work were the behavior of a charged particle sourced by quantum vacuum fluctuations of the electromagnetic field could be useful to probe space topology \cite{bessa2020a}.

Some comments on the use of smooth switching functions to describe transitions between physical states is here in order. First of all, it should be understood that such transitions may not occur instantaneously. The interval of time taken by a system to evolve between two different states may be dependent on parameters related to its dynamics, as for instance the approaching velocity between a reflective wall and a charged particle. Even backreaction phenomena behave like that, as for example, happens with diffraction radiation \cite{karlovets2008}.
In fact, switching functions are greatly related to the layout of an experiment and the corresponding measurement process. Typical divergences appearing in idealized scenarios where the interaction is instantaneously turned-on and -off, are naturally regularized when smooth transitions are implemented, suggesting that these functions bring more reality in the description of the physical system.

Recently, velocity fluctuations of a scalar charged test particle interacting with a real massive scalar field in the presence of a perfectly reflecting flat boundary were examined \cite{Camargo2019}, where the effects of the field mass and spacetime dimension were addressed. From an experimental perspective, this system serves as a toy model for feasible scenarios of electrically charged particles in the interior of two or three dimensional conductors, in which electromagnetic disturbances obey a dispersion relation in the form $\omega^2=c^2k^2+\omega_p^2$, where $\omega_p$ is the plasma frequency \cite{Sopova2002}. However, finite temperature effects were not considered, which are certainly more prominent in most of the measurements. In this work, we address the question of how to probe quantum fluctuations induced by an idealized boundary of a massive scalar field at finite temperature using velocity dispersions of test particles, following the smoothing measurement methodology. 
Novel consequences concerning the distance behavior of the dispersions are discussed. As the presence of mass in the field suppresses the thermal contribution, the dominance near the wall depends on it.

The paper is organized as follows. In the next section a brief review of a massive quantum scalar field at finite temperatures is presented, where important quantities for the present work are calculated. Then, in Sec.~\ref{secIII}, the model used for the smooth activated interaction of a test particle with the quantum field is introduced together with some assumptions and the chosen switching functions, which model the transition. The main results are given in Sec.~\ref{secIV}, where the behavior of the stochastic motion of the test particle is studied when it is immersed in a pure thermal bath (without a boundary), and when a reflecting boundary is added to the system. In this later scenario it is shown that negative dispersions of  the particle velocity occur. Hence, negative dispersions, already known to happen when only a boundary is present, not only survive but are enhanced when the system is allowed to experience a thermal bath. Therefore, as the purely thermal contribution is independent from the boundary one, the effects can be separated, either due to good knowledge of the pure thermal dispersion or by means of the physical setup. So, increasing the temperature facilitates the detection of subvacuum effects. Graphical analysis is provided for $D=2$ and $D=3$ space dimensions.  
The late-time regime of the dispersions is examined in Sec.~\ref{secV}, underlining the important role of the transition time between field physical states.  Finally, in Sec.~\ref{secVI}, the field is investigated near the wall, where the mass plays an important role concerning vacuum versus thermal dominance when the wall is approached. The ~\ref{appendix} contains the detailed calculation of the late time regime.

 Henceforth, units are such that $c=\hbar=k_{_{B}}=1$, where $k_{_{B}}$ is the Boltzmann constant. 
\section{Quantum aspects of the background field}
\label{secII}
In this section we present the relevant propagators needed in the following sections. We shall adopt the canonical quantization prescription to formally expand the quantum field in terms of normal modes, and this section is relevant to review important results concerning thermodynamic stability of the system under study. We start with a massive real scalar field $\phi(t,\textbf{x})$ in the $D+1$ dimensional Minkowski spacetime. The field dynamics is ruled by the Klein-Gordon equation $(\partial_t^2-\nabla^2+m^2)\phi(t,\textbf{x})=0$. We expand the field as
\begin{equation}
\phi(t,\textbf{x})=\int \frac{\mathrm{d}^{D}\textbf{k}}{\sqrt{2\omega(2\pi)^{D}}}\left[a_{\textbf{k}}\textrm{e}^{-i(\omega t-\textbf{k}\cdot\textbf{x})}+a^{\dagger}_{\textbf{k}}\textrm{e}^{i(\omega t-\textbf{k}\cdot\textbf{x})}\right],
 \nonumber 
\end{equation}
where $\omega=(\textbf{k}^2+m^2)^{1/2}$ and
the operators $a_{\textbf{k}}$ and $a^{\dagger}_{\textbf{k}}$ are such that $[a_{\textbf{k}},a^{\dagger}_{\textbf{k}'}]=\delta^{D}(\textbf{k}-\textbf{k}')$, with all other commutators vanishing.

The quantities of interest in the present work are obtained from the Hadamard two-point function at finite temperature $G^{(1)}_\beta(t,\textbf{x};t',\textbf{x}')=\langle\{\phi(t,\textbf{x}),\phi(t',\textbf{x}')\}\rangle_{\beta}$, where the braces denote the anticommutator between the fields, and the expectation value is taken in the grand canonical ensemble \cite{Davies1982}, through which the temperature $T=\beta^{-1}$ is introduced. 

This function can be written in terms of Bessel functions as
\begin{align}
G^{(1)}_{\beta}(t,\x;t',\x')=&G^{(1)}(t,\x;t',\x')\nonumber\\
+&\frac{2}{(2\pi)^\frac{D}{2}|\Delta\x|^{\frac{D}{2}-1}}
\textrm{Re}\sum_{l=1}^{\infty}\int_0^\infty\mathrm{d}k\frac{k}{\omega}\e^{-i\omega(\Delta t-i\beta l)}\nonumber\\
\times& k^{\frac{D}{2}-1}J_{\frac{D}{2}-1}(k|\Delta\x|),
\label{eq2}
\end{align}
where we denote $\Delta a=a-a'$, and $G^{(1)}(t,\x;t',\x')$ is the zero temperature Hadamard function
\begin{align}
G^{(1)}(t,\x;t',\x')=&\frac{1}{(2\pi)^\frac{D}{2}|\Delta\x|^{\frac{D}{2}-1}}\nonumber\\
&\times\textrm{Re}\int_0^\infty\mathrm{d}k\frac{k}{\omega}\e^{-i\omega\Delta t}k^{\frac{D}{2}-1}J_{\frac{D}{2}-1}(k|\Delta\x|).
\label{eq3}
\end{align}
Integrating Eq.  (\ref{eq2}) we find \cite{gradshteyn}
\begin{align}
G^{(1)}_{\beta}(t,\x;t',\x')&=\frac{1}{\pi}\textrm{Re}\left[\left(\frac{m}{2\pi i \sigma_0}\right)^{\frac{D-1}{2}}K_{\frac{D-1}{2}}(im\sigma_0)\right]\nonumber\\
&+\frac{2}{\pi}\textrm{Re}\left[\sum_{l=1}^{\infty}\left(\frac{m}{2\pi i \sigma_l}\right)^{\frac{D-1}{2}}K_{\frac{D-1}{2}}(im\sigma_l)\right]
\label{eq4}
\end{align}
with $\sigma_l=[(\Delta t-i\beta l)^2-(\Delta \x)^2]^{1/2}$. In integrating Eq.~(\ref{eq3}) we set $\Delta t=t-t'-i\epsilon$, $\epsilon$ being an infinitesimal positive number to ensure convergence when $l=0$, underlining the distributional character of the above function.

Finally, in order to find the renormalized finite temperature Hadamard function in the presence of an infinite Dirichlet's wall at $x_1=0$, where $\phi(t,x_1=0,x_2,...,x_D)=0$, we use the image method \cite{brown69}, which gives
\begin{align}
G^{(1)}&_{\beta,\textrm{Ren}}(t,\x;t',\x')=\nonumber
-\frac{1}{\pi}\textrm{Re}\left[\left(\frac{m}{2\pi i \sigma^{+}_0}\right)^{\frac{D-1}{2}}K_{\frac{D-1}{2}}(im\sigma^{+}_0)\right]\nonumber\\
&+\frac{2}{\pi}\textrm{Re}\sum_{l=1}^{\infty}\left[\left(\frac{m}{2\pi i \sigma_l}\right)^{\frac{D-1}{2}}K_{\frac{D-1}{2}}(im\sigma_l)-(\sigma_l\leftrightarrow\sigma^{+}_l)\right]. 
\label{eq5}
\end{align}
Here $\sigma^{+}_l=[(\Delta t-i\beta l)^2-(\hat{\Delta}\x)^2]^{1/2}$, $\hat{\Delta}\x$ being $\Delta\x$ when $x'_1\rightarrow -x'_1$. Moreover, the function was already renormalized, i.e., the free vacuum term was subtracted, as it has no contribution to the stochastic motion \cite{johnson2002}.

Note that the above expression is conveniently divided into three contributions: the first one on the right hand side is the modified vacuum term, due to the presence of boundary; the second one is the thermal contribution; and the last one is the mixed contribution, from the modified vacuum at finite temperature. Such division will be used when treating the dispersions.

A few remarks about the massive scalar field at finite temperature are imperative. Firstly, as the retarded propagator is constructed through the Pauli-Jordan two-point function $G(t,\x;t',\x')=\left[\phi(t,\textbf{x}),\phi(t',\textbf{x}')\right]$, a c-number, it remains the same as the zero temperature case, so that the discussion in Ref. \cite{Camargo2019} holds. In particular, it was shown that the non-Huygensian character of the massive fields, and of the massless fields when $D$ is even, allowing signal to propagate with any velocity lower than the light, may induce peculiar behavior in experiments measuring nonlocal observables; such as oscillations in the Hadamard function. This is physically reasonable, as we do not expect the causal structure of a field to be modified by the presence of the heat reservoir. Secondly, at finite temperature, when $m=0$ and $D=2$, the system is out of equilibrium \cite{fulling}. This fact can be easily deduced by using the propagator presented in Eq.~\eqref{eq5}. In fact, recall that field fluctuations are measured by $\langle\phi^2\rangle_{\beta,{\rm Ren}}=(1/2)G^{(1)}_{\beta,{\rm Ren}}(t,\x;t,\x)$, which, as anticipated, can be conveniently written as the sum
\begin{equation}
\langle\phi^2\rangle_{\beta,{\rm Ren}}=\langle\phi^2\rangle_{\beta,{\rm vacuum}}+\langle\phi^2\rangle_{\beta,{\rm thermal}}+\langle\phi^2\rangle_{\beta,{\rm mixed}}\label{eqphi2}.
\end{equation}
We are interested in the pure thermal term, $\langle\phi^2\rangle_{\beta,{\rm thermal}}$, that is always present despite the presence of the boundary. In the limit of vanishing field mass, it can be shown that
\begin{equation}
\langle\phi^2\rangle_{\beta,{\rm thermal}}=\frac{1}{2\beta^{D-1}\pi^{(D+1)/2}}\Gamma\left(\frac{D-1}{2}\right)\sum_{l=1}^{\infty}\frac{1}{l^{D-1}},
\nonumber
\end{equation}
which is clearly divergent for smaller dimensions, characterizing thermodynamic instability. A detailed study of the local behavior of a scalar gas near a Dirichlet's wall was presented in Ref. \cite{delorenci2015}, where the authors used the Feynman propagator to calculate the observables. Note that the same analysis can also be done using the Hadamard function of Eq.~\eqref{eq5}.

\section{The interacting model}
\label{secIII}
Here a test point particle is used to probe fluctuations of the massive scalar field in $D+1$ dimensions that is in thermal equilibrium with some reservoir. It is assumed the regime of small velocities and velocities fluctuations, which justify the nonrelativistic treatment for the system. Thus, the interaction of the test particle with the field is given through the usual Newtonian force law
\begin{equation}
\frac{dv_i}{dt}=-g\frac{\partial\phi}{\partial x_i},
\nonumber 
\end{equation}
where $g$ is the charge to mass ratio of the particle. In this model we assume that the particle position does not change significantly and that backreaction effects can be neglected \cite{ford2004,delorenci2016}. Furthermore, a key ingredient in this analysis is that the particle interaction with the relevant field state is smoothly switched on and off (after a given measuring time $\tau$). 
This switching mechanism, which models the transition between physical states of the background field, is implemented by the so-called switching function $F(t)$, in such a way that the $i$-th component of the particle velocity can be integrated as
\begin{equation}
v_i(\tau)=-g\int_{-\infty}^{\infty}\mathrm{d}tF(t)\frac{\partial\phi(t,\x)}{\partial x_i}.
\nonumber 
\end{equation}

Thus one can see that fluctuations of the background field induce a stochastic force in the particle. As the field contributions are at thermal equilibrium, its mean expectation value is zero, and thus $\langle v_i \rangle_{\beta}=0$. However, it will induce dispersions of the particle velocity $\langle(\Delta v_i)^2\rangle_{\beta}=\langle v^2_i\rangle_{\beta}$. Indeed, even in the presence of a classical force, only quantum fluctuations will contribute to the dispersions along the classical trajectory. Hence, symmetrizing the product of the field to avoid ambiguities in the quantization, it follows that the velocity dispersions are given by
\begin{multline}
\braket{(\Delta v_i)^2}=\frac{g^2}{2}\left[\frac{\partial\ }{\partial x_i}\frac{\partial\ }{\partial x_i'}\int_{-\infty}^{\infty}F(t)\int_{-\infty}^{\infty}F(t')\right.\\           \times G^{(1)}_{\beta,{\rm Ren}}(t,\x;t',\x')\mathrm{d}t\,\mathrm{d}t' \bigg]_{\x'=\x}.
\label{eq9}
\end{multline}
Following the natural splitting suggested in Eq.~\eqref{eq4}, it is convenient to present the dispersions as in Eq.~\eqref{eqphi2}, i.e.,
\begin{align}
\braket{(\Delta v_i)^2}=&\braket{(\Delta v_i)^2}_{\textrm{vacuum}}+\braket{(\Delta v_i)^2}_{\textrm{thermal}}+\braket{(\Delta v_i)^2}_{\textrm{mixed}},
\label{eq10}
\end{align}
so that each contribution can be easily identified.

An useful physical intuition on the model comes from the setup proposed in \cite{delorenci2016}, where electrons are shot in the vacuum, passing parallel to a reflective plate (the contribution of the path of the particle is of subleading order) and being detected in a condensation plate. The fluctuations in the velocities when they leave the plate are measured as dispersions on the particles final position.

Moreover, it should be stressed that previous investigations concerning the case of a sudden transition showed that divergences appear for the vacuum contribution near the wall and when the interaction time equals two times the distance from the wall. Such divergences are due to over idealizations of the models and a more realistic treatment, the smooth switching, is able to resolve the UV divergences. In what follows, the function $F$ is assumed to be analytic. Notice, however, that actual measurements can only be modeled by functions of compact support, that cannot be analytic. For our choices of smooth functions, it can be shown by numeric simulations that the error committed by this approximation is small, and the advantage of choosing analytic smooth functions is that they lead to closed expressions for the dispersions in some cases.

\subsection{Smooth switchings}

\begin{figure}[h!]
\center
\includegraphics[width=0.45\textwidth]{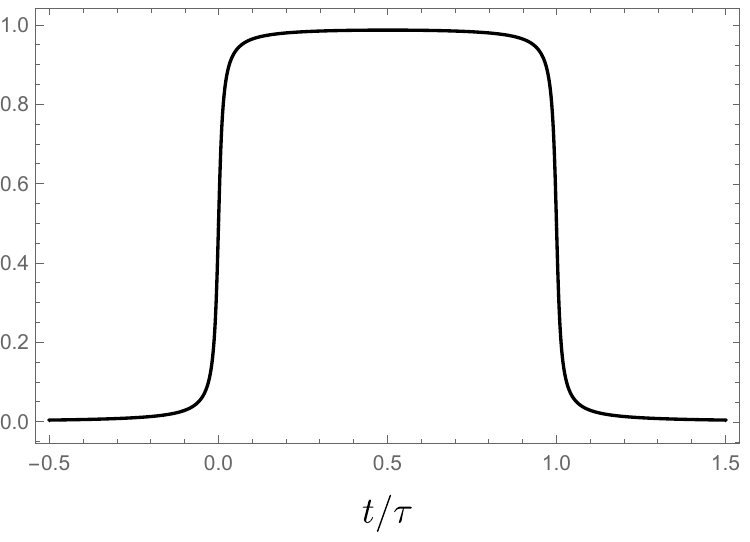}
\caption{Characteristic profile of the switching functions. Here, we plotted $F_{\tau_s,\tau}^{(2)}(t/\tau)$ for $\tau_s/\tau=0.01$.}
\label{figsf}
\end{figure}

In \cite{Camargo2019}, it was shown that when some interaction is smoothly switched on and off, high energy modes, that would otherwise imprint on the observables, are filtered out. In particular, this process is enough to regularize UV divergences. A straightforward choice for the switching is the one that models sudden processes, $F^{(0)}_{\tau}(t)=\Theta(t)\Theta(\tau-t)$, where $\tau$ is the measuring time. Notice that in this case, the switching time, hereafter denoted by $\tau_{s}$, is zero. If $F$ is to model a smooth switching, it must be normalized to the measuring time $\tau$, that is,
\begin{equation}
\int_{-\infty}^{\infty}\mathrm{d} tF(t)=\tau,
\nonumber 
\end{equation}
and it should recover the idealized sudden switching in some regime. A possible choice for such function is the generalized Lorentzian distribution $F^{(1)}_{n,\tau}(t)=c_{n}/[1+(2t/\tau)^{2n}]$, where $c_{n}=(2n/\pi)\sin (\pi/2n)$ \cite{delorenci2016,Barton1991A,Barton1991B}, which goes to $F^{(0)}_{\tau}$ as $n\rightarrow\infty$ and have the Fourier transform [we define the Fourier transform as $\hat{F}^{(1)}_{n,\tau}(\omega)=\int\mathrm{d} t\textrm{e}^{-i\omega t}F^{(1)}_{n,\tau}(t)$] 
\cite{Camargo2019}
\begin{equation}
    \hat{F}^{(1)}_{n,\tau}(\omega)=\frac{i\tau\pi c_n}{2n}\sum_{q=n}^{2n-1}\psi_{n,q}\,\e^{-i\omega \tau \psi_{n,q}/2},
    \label{eq11}
\end{equation}
with $ \psi_{n,p}=\exp[i(\pi/2n)(1+2p)] $. Nonetheless, the above choice of switching function is such that the transition time is proportional to the interaction time, so that for $\tau\rightarrow \infty $ the transition does never occurs and for $\tau\rightarrow 0$, it behaves like the sudden switching. Hence, it is useful to introduce another switching function for which the switching time $\tau_s$ is detached from the interaction time, for that we choose \cite{Camargo2018}
\begin{equation}
F_{\tau_s,\tau}^{(2)}(t)=\frac{1}{\pi}\left[\arctan \left(\frac{t}{\tau_s}\right)+\arctan \left(\frac{\tau-t}{\tau_s}\right)\right],
\label{eq12}
\end{equation} 
which also gives the sudden transition as $\tau_s\rightarrow 0$, and its Fourier transform is $\hat{F}^{(2)}_{\tau_s,\tau}(\omega)=(1/i\omega)(1-\e^{-i\omega\tau})\e^{-\tau_s|\omega|}$. The generic profile of these switching functions is depicted in Fig.~\ref{figsf}.

It can be seen from the Fourier transforms $\hat{F}^{(2)}_{\tau_s,\tau}(\omega)$ and $\hat{F}^{(1)}_{n,\tau}(\omega)$ that the physical mechanism behind the regularization of UV divergences by these switching functions is an extra exponential decay in the frequency space.

At last, the consequences of different choices of $n$ and $\tau_s$, together with a comparison between different choices of switching functions, were already investigated in earlier publications \cite{delorenci2016,delorenci2019b,Camargo2019}. As it was seen, once one have specified the switching time $\tau_s$, which is related to $n$ in $F^{(1)}_{n,\tau}(t)$, and the interaction time $\tau$, the physical system and the corresponding observables become completely characterized, independent of the choice of switching function.

\section{Velocity dispersions induced by the field fluctuations}
\label{secIV}

We now start the presentation of the main results of this work. As discussed in the introduction, we separate the contributions to the dispersions in two parts, one that comes from the thermal bath only, and exists regardless of the boundary, and the additional effects appearing when the boundary is present. As we shall see, the scalar field boundary-modified vacuum state is such that when in presence of a heat reservoir, unexpected effects appear. We start by studying the ever present thermal-induced dispersions.
In what follows, we shall use the superscripts $(1)$ or $(2)$ to indicate the choice of switching defined by Eqs. (\ref{eq11}) and (\ref{eq12}), respectively, and a subscript $D$ indicating the spatial dimension.

\subsection{No boundary thermal dispersions}
\label{secIV.A}

The pure thermal contribution is obtained by substituting Eq.~\eqref{eq5}, apart from the boundary contribution, in Eq.~\eqref{eq9}, which results in
\begin{multline}
	\langle(\Delta v_i)^2\rangle_{D,\textrm{thermal}}
        =g^2\lim_{\x\rightarrow \x'}\left[\frac{\partial}{\partial x_i}\frac{\partial}{\partial x'_i}\frac{1}{(2\pi)^{\frac{D}{2}}|\Delta \x|^{\frac{D}{2}-1}}\right.\\ \left.\times\sum_{l=1}^{\infty}\int_{0}^{\infty}\mathrm{d} k\frac{k^{\frac{D}{2}}}{\omega}|\hat{F}(\omega)|^2e^{-l\beta\omega}J_{\frac{D}{2}-1}(k|\Delta \x|)\right].
    \label{eq13}
\end{multline}
Using the sample function $F^{(1)}_{n,\tau}(t)$ with Fourier transform given by Eq. \eqref{eq11}, the above expression can be integrated as
\begin{multline}
\langle(\Delta v_i)^2\rangle^{(1)}_{D,\textrm{thermal}}=\frac{2g^2}{\beta^{D-1}}\left[\frac{(\tau/\beta)\pi c_n}{2n}\right]^2\sum_{p,q=n}^{2n-1}\psi_{n,p}\psi^{*}_{n,q}\\\times\sum_{l=1}^{\infty}\left(\frac{m\beta}{2\pi\sqrt{-a^2_l}}\right)^{\frac{D+1}{2}}K_{\frac{D+1}{2}}\left(m\beta\sqrt{-a^2_l}\right),
\label{eq14}
\end{multline}
with $a_l=(\tau/2\beta)(\psi_{n,p}-\psi^{*}_{n,q})-i\,l$, and the sub-index $i$ appearing in $(\Delta v_i)^2$ can be any of the components, as expected from the isotropic nature of the thermal bath. Notice that the dispersions are homogeneous, and are exponentially suppressed as $\beta\rightarrow\infty$, which is the case for free vacuum.

\begin{figure}[h!]
\center
\includegraphics[width=0.45\textwidth]{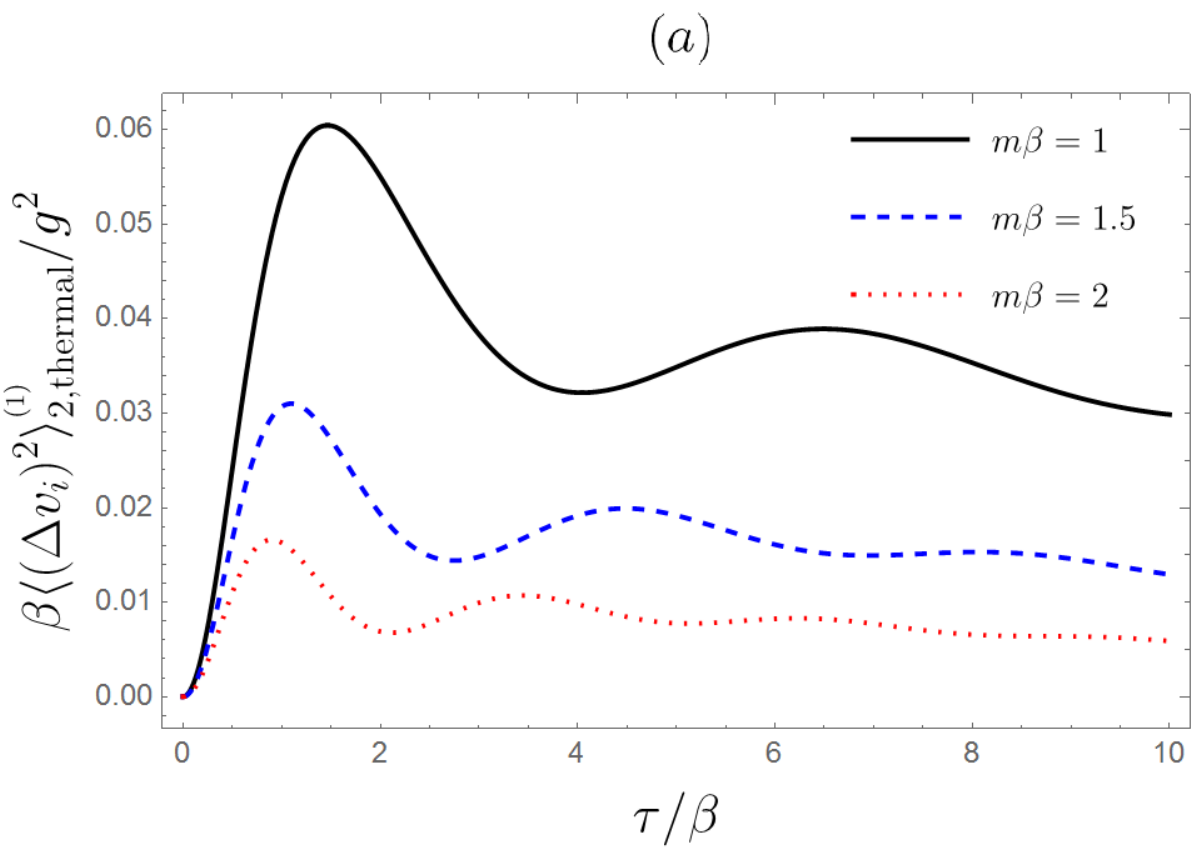}
\includegraphics[width=0.45\textwidth]{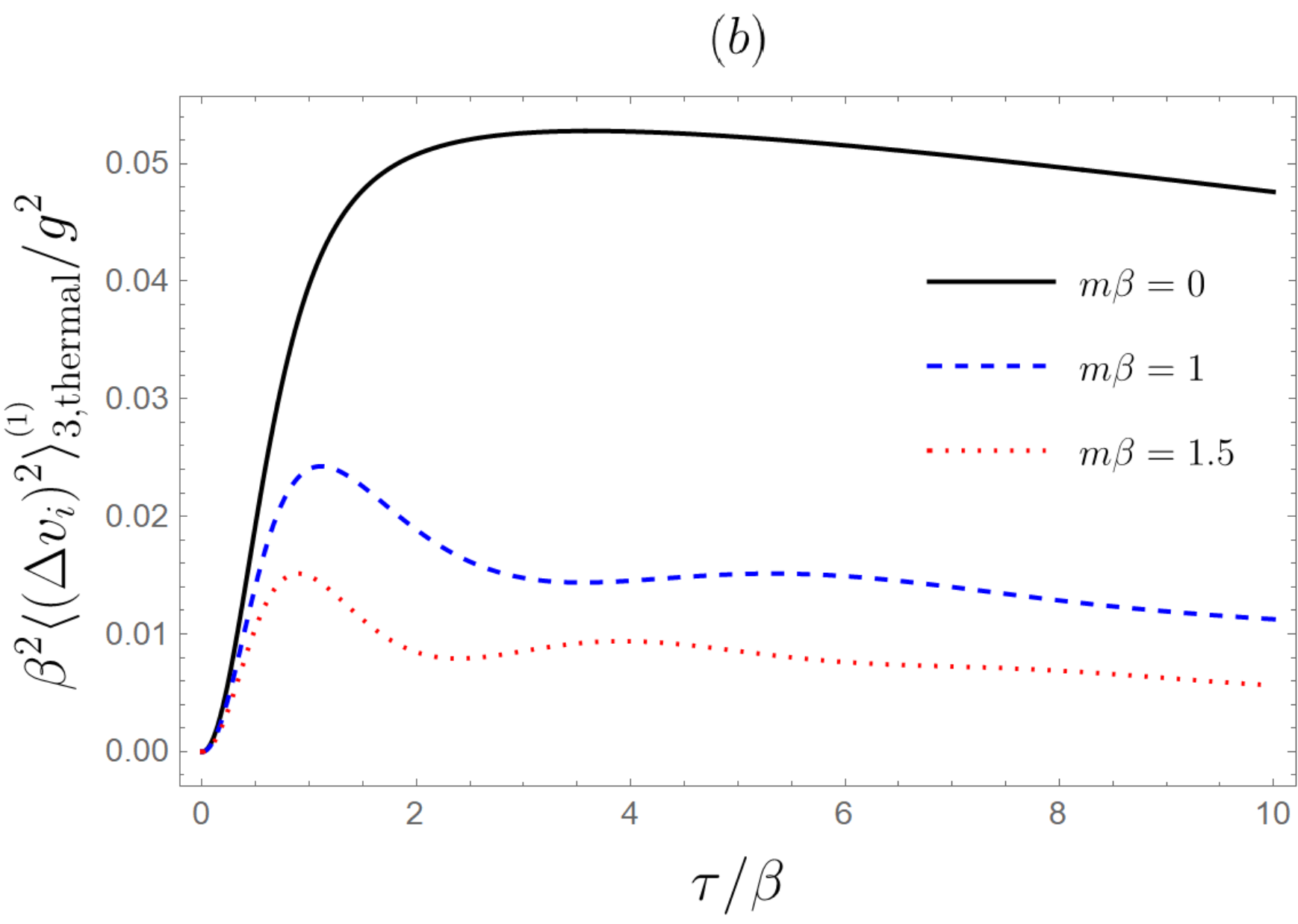}
\caption{Velocity dispersions caused by a thermal bath for different values of the mass with fixed $n=20$. (a) Dispersions for $D=2$. (b) Dispersions for $D=3$. As for $D=3$ the massless limit is well-behaved, no oscillations appear in this case.}
\label{fig2}
\end{figure}  

The dispersions are depicted in Fig.~\ref{fig2}. One can see that their behavior does not change much for different values of $D$, being just weaker as $D$ increases. Also, for the massless field in $D=3$, which is infrared well-behaved, the dispersions do not oscillate, as expected. Notice that they depend on $m\beta=m/T$ and thus there is an interplay between the field mass and the system temperature. As the mass grows the field acquires more inertia and therefore reacts less to the thermal energy, whereas as the temperature increases, the oscillatory pattern is suppressed. 

Recall that when the field mass goes to zero and $D=2$, the system is in the vicinity of a phase transition. However, the velocity dispersion is still defined in this scenario, and we find that
\begin{align}
\langle(\Delta v_{i})^2\rangle^{(1)}_{2,\textrm{thermal}}\stackrel{m\rightarrow0}{=}&-\frac{\pi}{\beta}\left(\frac{g\tau c_n}{4\beta n}\right)^2\nonumber\\
&\times\sum_{p,q=n}^{2n-1}\psi_{n,p}\psi^{*}_{n,q}\psi^{(2)}(1+ia_0),
\label{eq16}
\end{align}
where $\psi^{(2)}$ is a Polygamma function \cite{gradshteyn}.

Notice that for $D=2$, $\braket{(\Delta v_{i})^2}^{(1)}_{2,\textrm{thermal}}\rightarrow 0$ as $\tau\rightarrow\infty$, which is an apparent regularization. However, as we have seen, this choice of sample function is not well suited to study the late-time regime, as then the transition time goes to infinity as well. Hence, this dispersion must be addressed with another choice of sample function.

\subsection{Boundary-induced dispersions}
\label{secIV.B}

The pure thermal contribution to the velocity dispersions described by Eq. (\ref{eq14}) does not  change when a boundary is added to the system. However, two other contributions will be activated by the presence of the boundary, which are those anticipated in Eq. (\ref{eq10}). Thus, the boundary effects can in principle be detected separately, either by removing the well known pure thermal contribution in the final outcomes or by means of a setup arrangement -- the particle could be already immersed in the scalar gas when the wall is placed. In such cases the thermal contribution would be just a constant late-time value, which can be made small for switching times long enough. Henceforth, we now investigate the boundary contributions to the dispersions, which consist of the modified vacuum contribution and a mixed term. The former one was investigated in a previous work \cite{Camargo2019}, while the later unveils  the important interplay between boundary and thermal effects. Hence, we define 
\begin{equation}
\braket{(\Delta v_i)^2}_{\textrm{boundary}}=\braket{(\Delta v_i)^2}_{\textrm{vacuum}}+\braket{(\Delta v_i)^2}_{\textrm{mixed}}.
\nonumber 
\end{equation}
The calculations of the above dispersions are analogous as for the thermal case, but with $\Delta \x$ replaced by $\hat{\Delta} \x$. The introduction of the reflecting boundary breaks the isotropy of the velocity fluctuations, with the perpendicular direction to it being different from the others. 
\begin{figure}[h!]
\center
\includegraphics[width=0.45\textwidth]{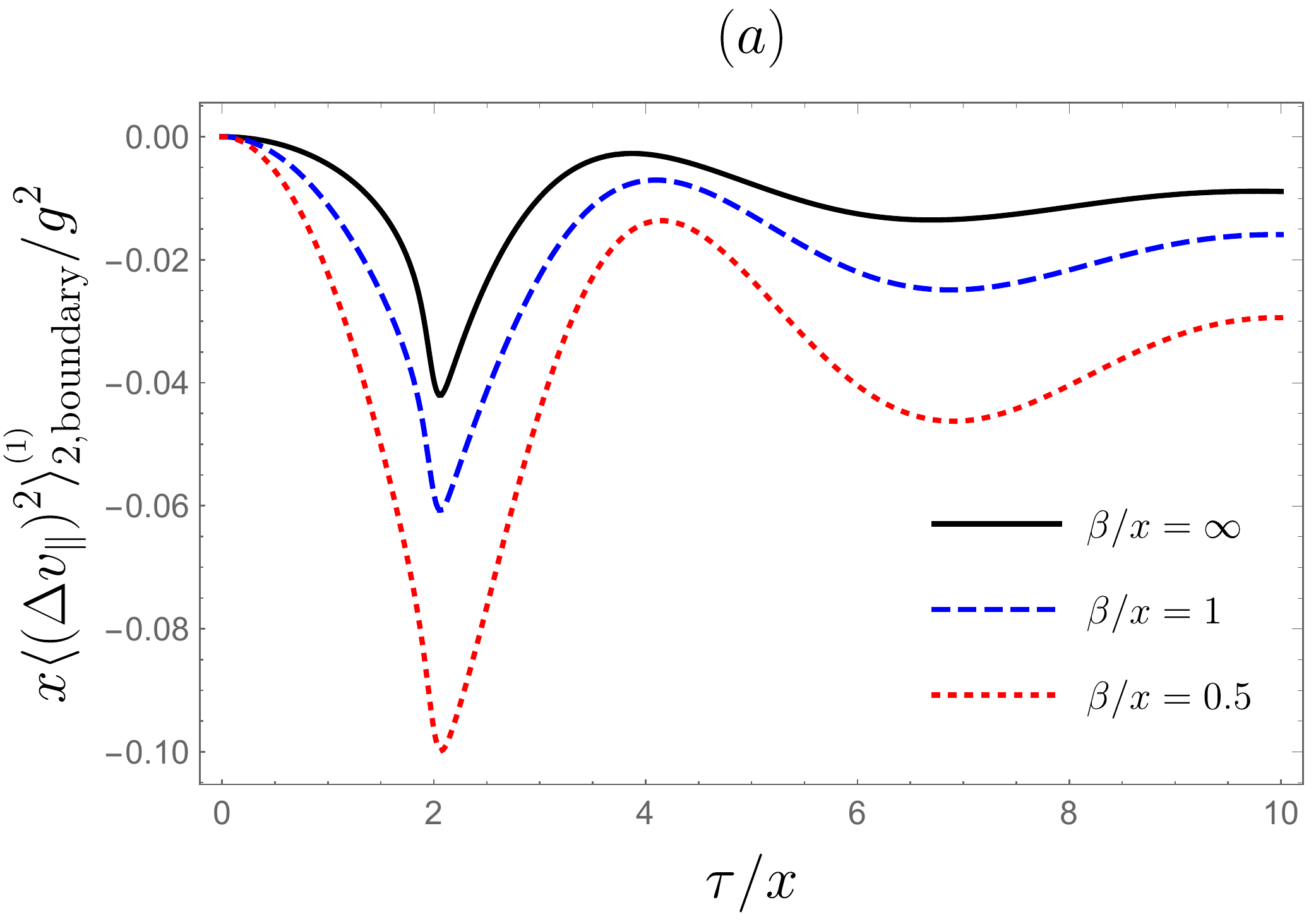}
\includegraphics[width=0.45\textwidth]{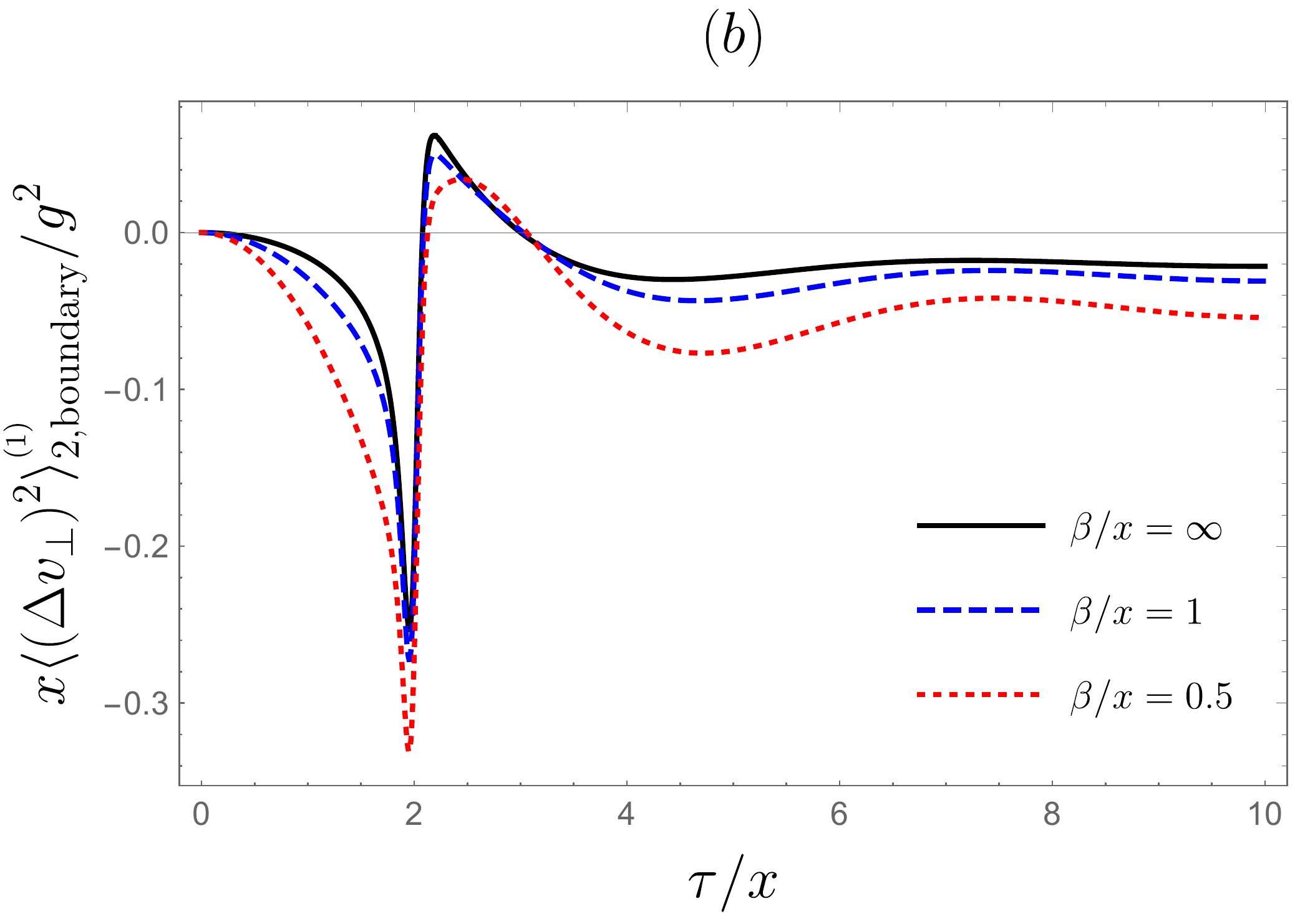}
\caption{Dispersion caused by a reflective boundary at finite temperature for $D=2$. We set $n=20$ and $mx=1$. (a) Velocity dispersions in the parallel direction. (b)  Velocity dispersions in the perpendicular direction. It can be seen that the magnitude of the dispersions increases with the temperature, thus enhancing a subvacuum behavior.}
\label{fig3}
\end{figure}  
\begin{figure}[h!]
\center
\includegraphics[width=0.45\textwidth]{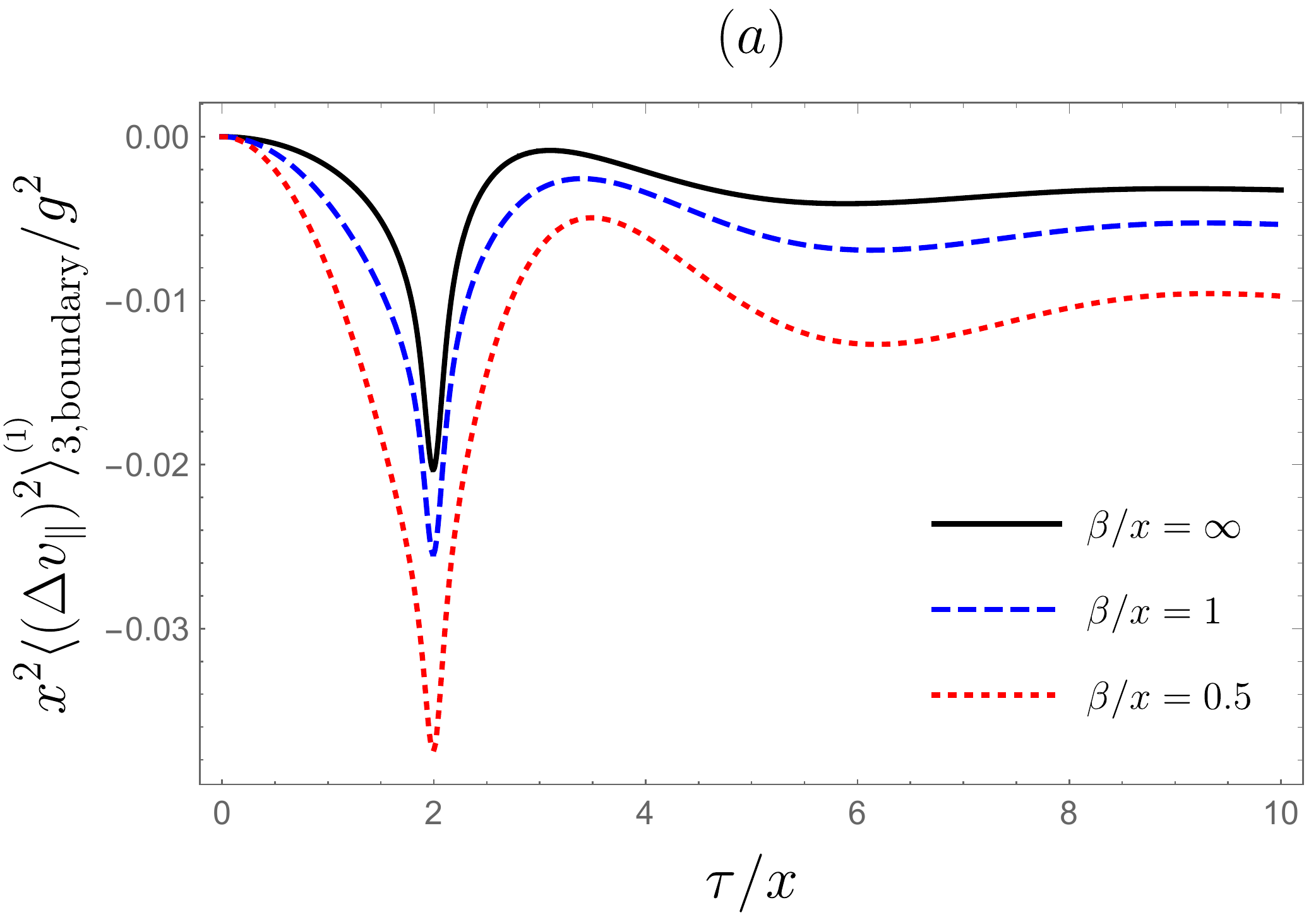}
\includegraphics[width=0.45\textwidth]{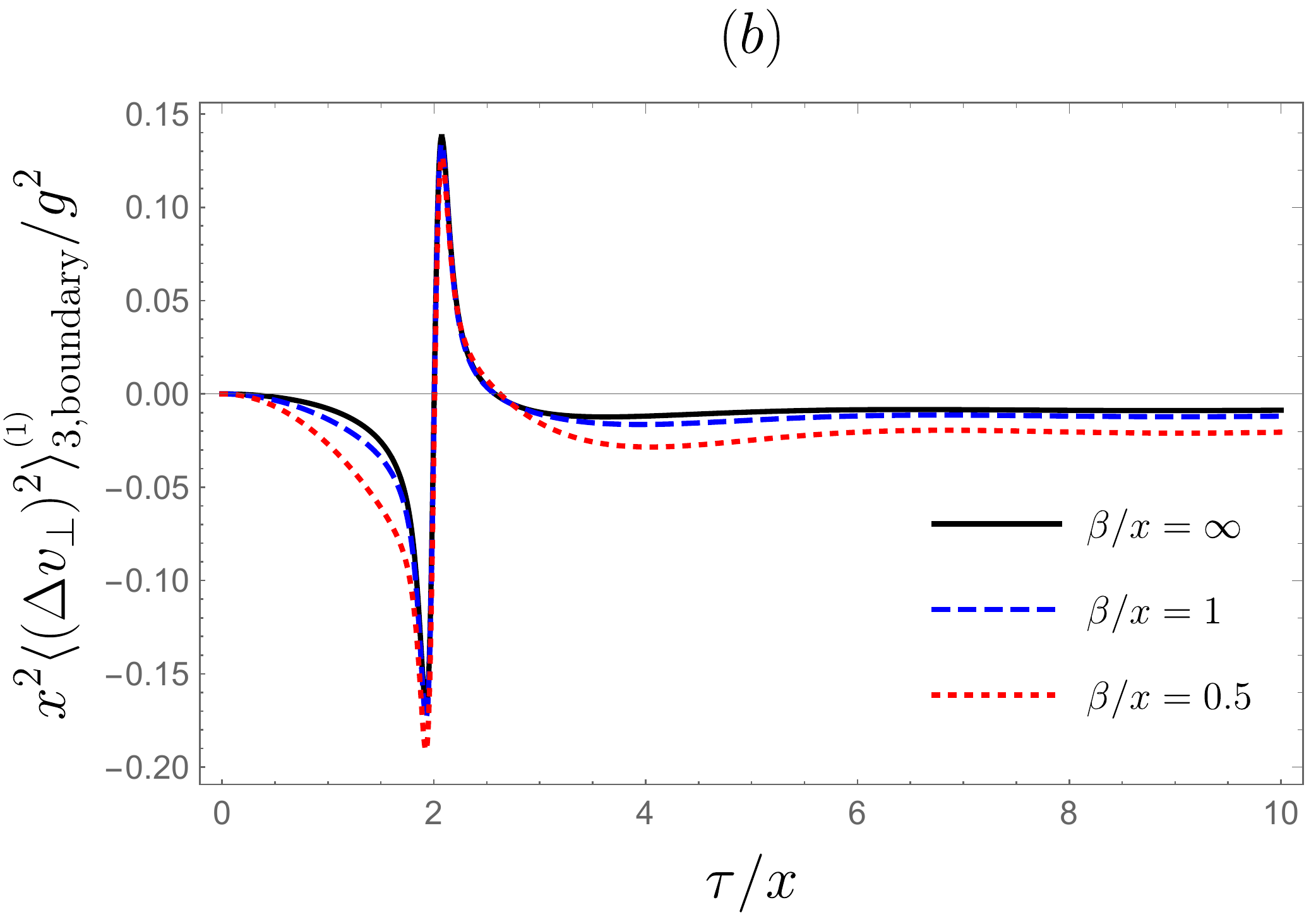}
\caption{Dispersion caused by a reflective boundary at finite temperature for $D=3$. We set $n=20$ and $mx=1$. (a) Velocity dispersions in the parallel directions. (b)  Velocity dispersions in the perpendicular direction. The temperature effects are lower as compared to $D=2$ case.}
\label{fig4}
\end{figure}  

In this scenario the mixed contribution to the dispersions are given by, 
\begin{align}
&\langle(\Delta v_\parallel)^2\rangle^{(1)}_{D,\textrm{mixed}}=\frac{-2g^2}{x^{^{D-1}}}\left[\frac{(\tau/x)\pi c_n}{2n}\right]^{2}\sum_{p,q=n}^{2n-1}\psi_{n,p}\psi^{*}_{n,q}\nonumber\\
&\times\sum_{l=1}^{\infty}\left(\frac{mx}{4\pi\sqrt{1-\gamma^2_{l}}}\right)^{\frac{D+1}{2}}K_{\frac{D+1}{2}}\left(2mx\sqrt{1-\gamma^2_{l}}\right),
\nonumber 
\\
&\langle(\Delta v_\perp)^2\rangle^{(1)}_{D,\textrm{mixed}} =-\langle(\Delta v_\parallel)^2\rangle^{(1)}_{D,\textrm{mixed}} \nonumber\\
&\hspace{3cm} +8\pi x^2\langle(\Delta v_\parallel)^2\rangle^{(1)}_{D+2,\textrm{mixed}},
 \nonumber 
\end{align}
where we have defined $\gamma_l= (\beta/2x)a_l$,  $x$ being the distance to the boundary.

Direct inspection of the propagator in Eq.~\eqref{eq5} shows that the vacuum contribution to the dispersions is just half the mixed part with $l=0$, so that
\begin{align}
    &\langle(\Delta v_\parallel)^2\rangle^{(1)}_{D,\textrm{vacuum}}=\frac{-g^2}{x^{D-1}}\left[\frac{(\tau/x)\pi c_n}{2n}\right]^2\sum_{p,q=n}^{2n-1}\psi_{n,p}\psi^{*}_{n,q}\nonumber\\
		&\times\left(\frac{mx}{4\pi\sqrt{1-\gamma^2_0}}\right)^{\frac{D+1}{2}}K_{\frac{D+1}{2}}\left(2mx\sqrt{1-\gamma^2_0}\right),
\nonumber 
\\
    &\langle(\Delta v_\perp)^2\rangle^{(1)}_{D,\textrm{vacuum}}=-\langle(\Delta v_\parallel)^2\rangle^{(1)}_{D,\textrm{vacuum}}\nonumber\\
		&\hspace{3cm} +8\pi x^2\langle(\Delta v_\parallel)^2\rangle^{(1)}_{D+2,\textrm{vacuum}}.
 \nonumber 
\end{align}

Just as the thermal contributions, the mixed contributions also get suppressed when $\beta\rightarrow\infty$ and only the modified vacuum is left, recovering previous results \cite{Camargo2019}.
\begin{figure}[h!]
\center
\includegraphics[width=0.45\textwidth]{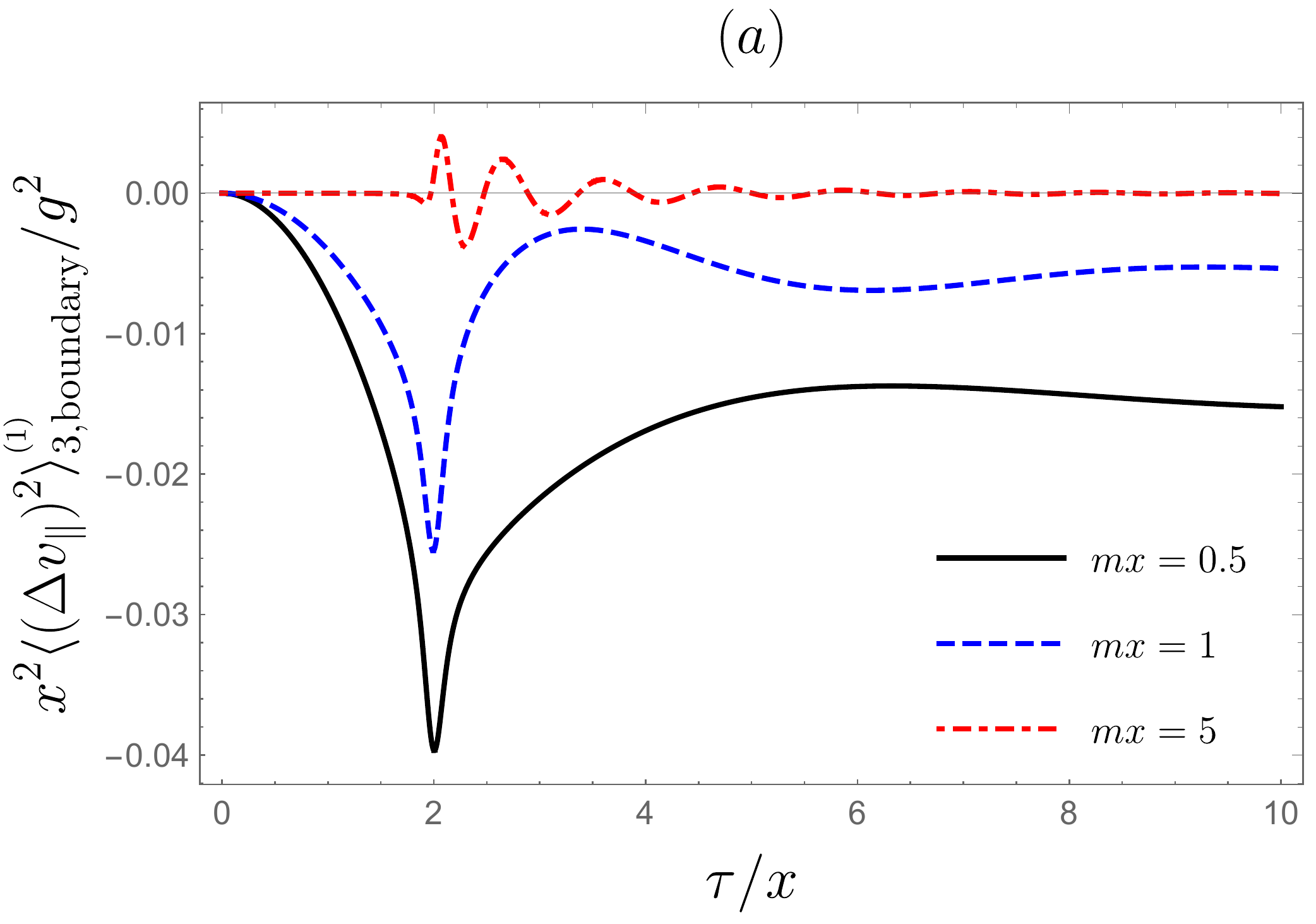}
\includegraphics[width=0.45\textwidth]{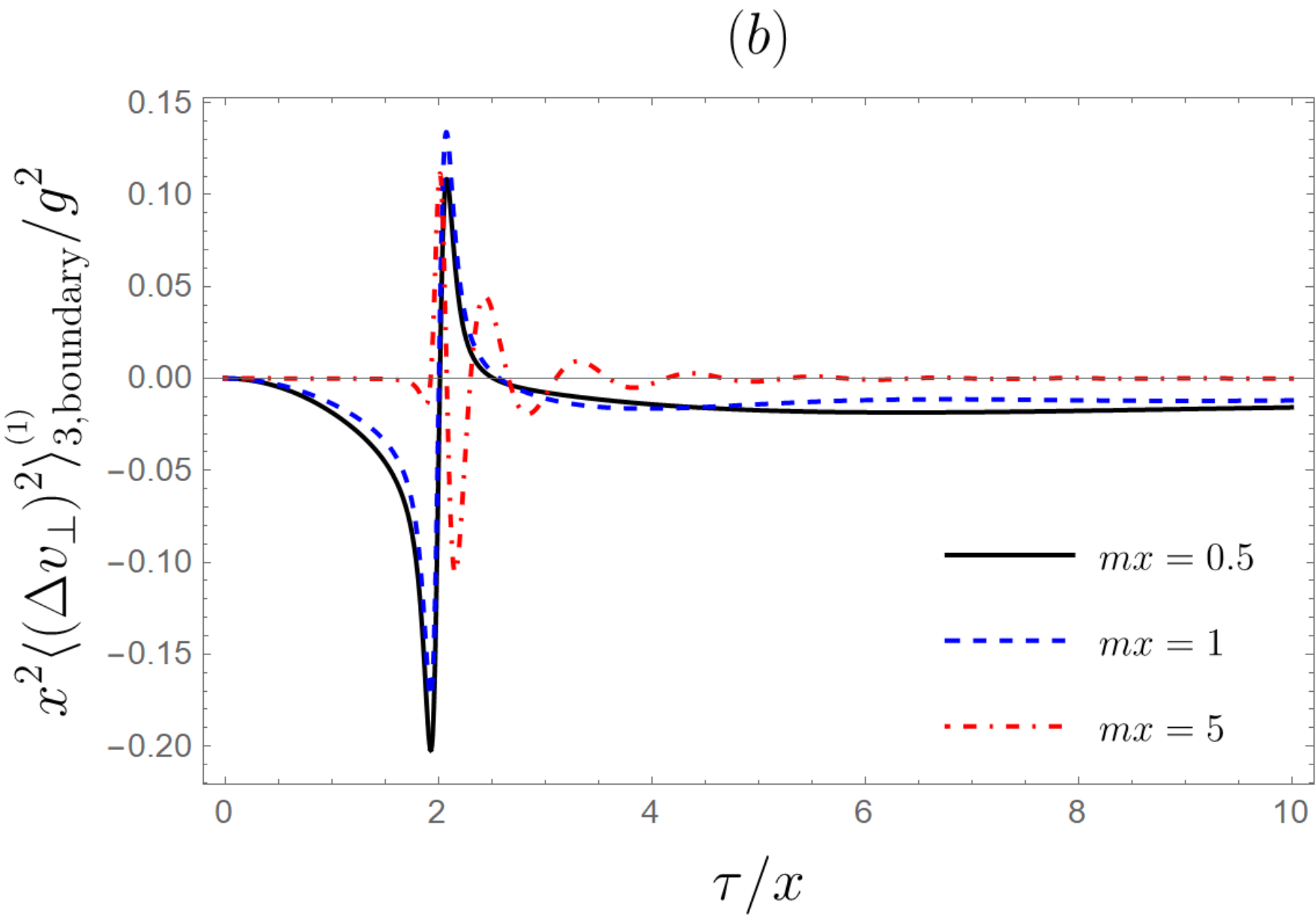}
\caption{Dispersion caused by a reflective boundary at finite temperature for $D=3$. We set $n=20$ and $\beta/x=1$. (a) Velocity dispersions in the parallel directions. (b)  Velocity dispersions in the perpendicular direction. As the field mass increases, the modified vacuum dominates over the finite temperature contribution.}
\label{fig5}
\end{figure} 

Figures \ref{fig3} and \ref{fig4} depict the time behavior of boundary dispersions $\braket{(\Delta v_i)^2}_{\textrm{boundary}}$ for $D=2$ and $3$. These figures unveil an intriguing result,  that the interplay between boundary and temperature increases the magnitude of the subvacuum effect. On the other hand, purely thermal contribution, studied in the last section, goes the opposite way. The net effect will be that the total dispersion will still exhibit subvacuum behavior for a certain range of parameters, as it will be discussed in the next section. 
Notice that characteristic peaks appear about $\tau=2x$, which amounts to the time interval taken by a high energy scalar wave for a round trip between the charged particle and the boundary \cite{ford2004,delorenci2014,delorenci2016}. Their magnitudes and widths are dependent on the duration of the transition, as expected. Remember that a divergence appear at this very same point when an idealized sudden transition is implemented.

The behavior of the dispersions for some convenient values of $mx$ is present in Fig.~\ref{fig5}. Notice that as $m$ takes larger values, the mixed contributions get suppressed and the vacuum term dominates, recovering the zero temperature case.

Closing this section, it should stressed that the dispersions were here obtained by using a sample function which describes a switching time that depends on the measuring time $\tau$. So, in this case a late-time regime would also be a regime of long-lasting switching. Hence, if we wish to satisfactorily study the dispersions at late-time regime the sample function defined in Eq.~(\ref{eq12}) should be considered. This is the subject of the next section.

\section{Late-time behavior of the velocity dispersions}
\label{secV}

The stochastic motion induced by quantum field fluctuations crucially depends on the switching time $\tau_s$ between vacuum states, and this aspect already shows a difference when compared to a Brownian motion. In order to investigate the influence of  $\tau_s$ on the dispersions we may calculate them by using the switching function $F_{\tau_s,\tau}^{(2)}$, for which $\tau_s$ is an independent parameter, thus properly allowing the description of a late-time regime. Detailed calculations of the dispersions examined in this section are given in \ref{appendix}. 

\begin{figure}[h!]
\center
\includegraphics[width=0.48\textwidth]{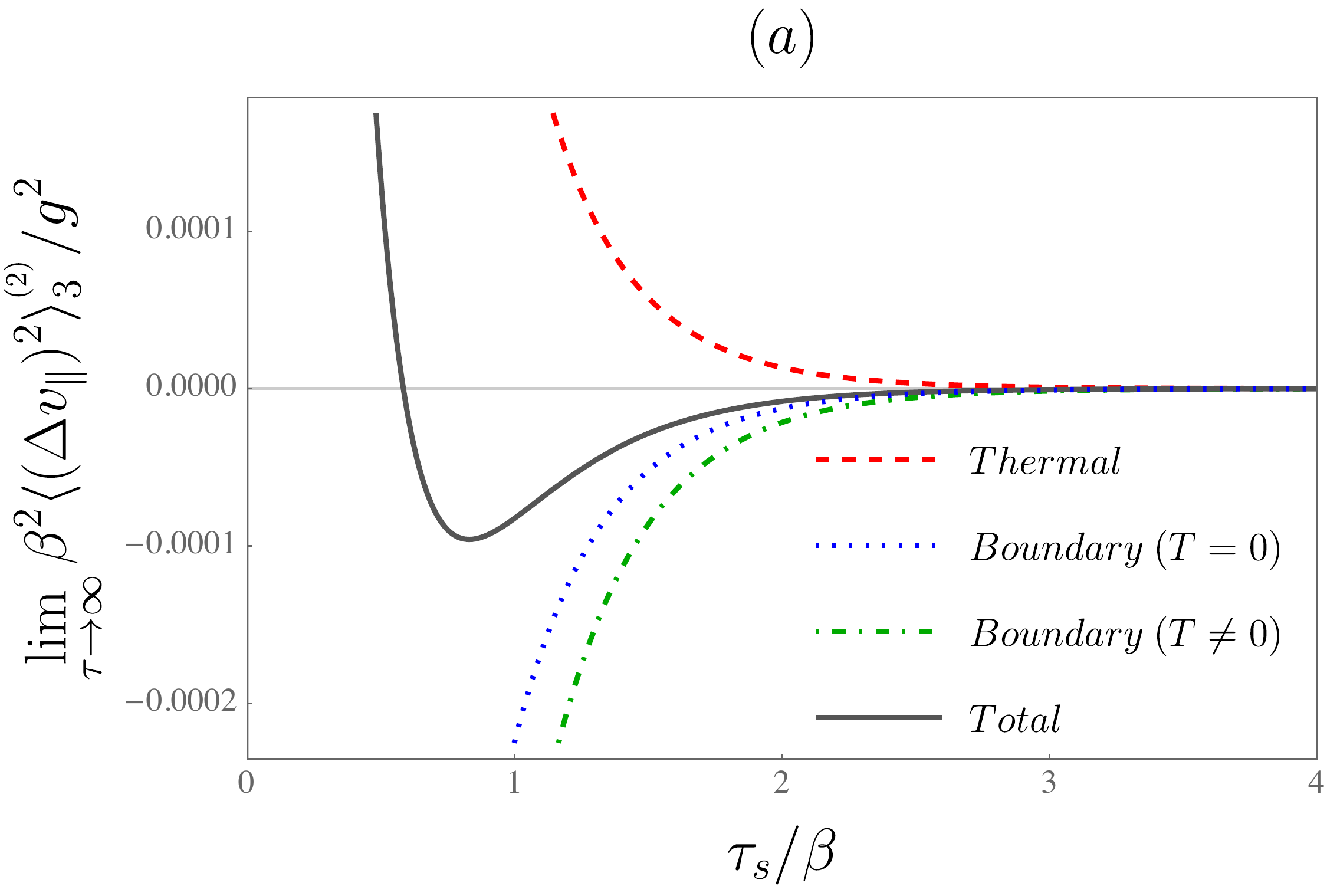}
\includegraphics[width=0.48\textwidth]{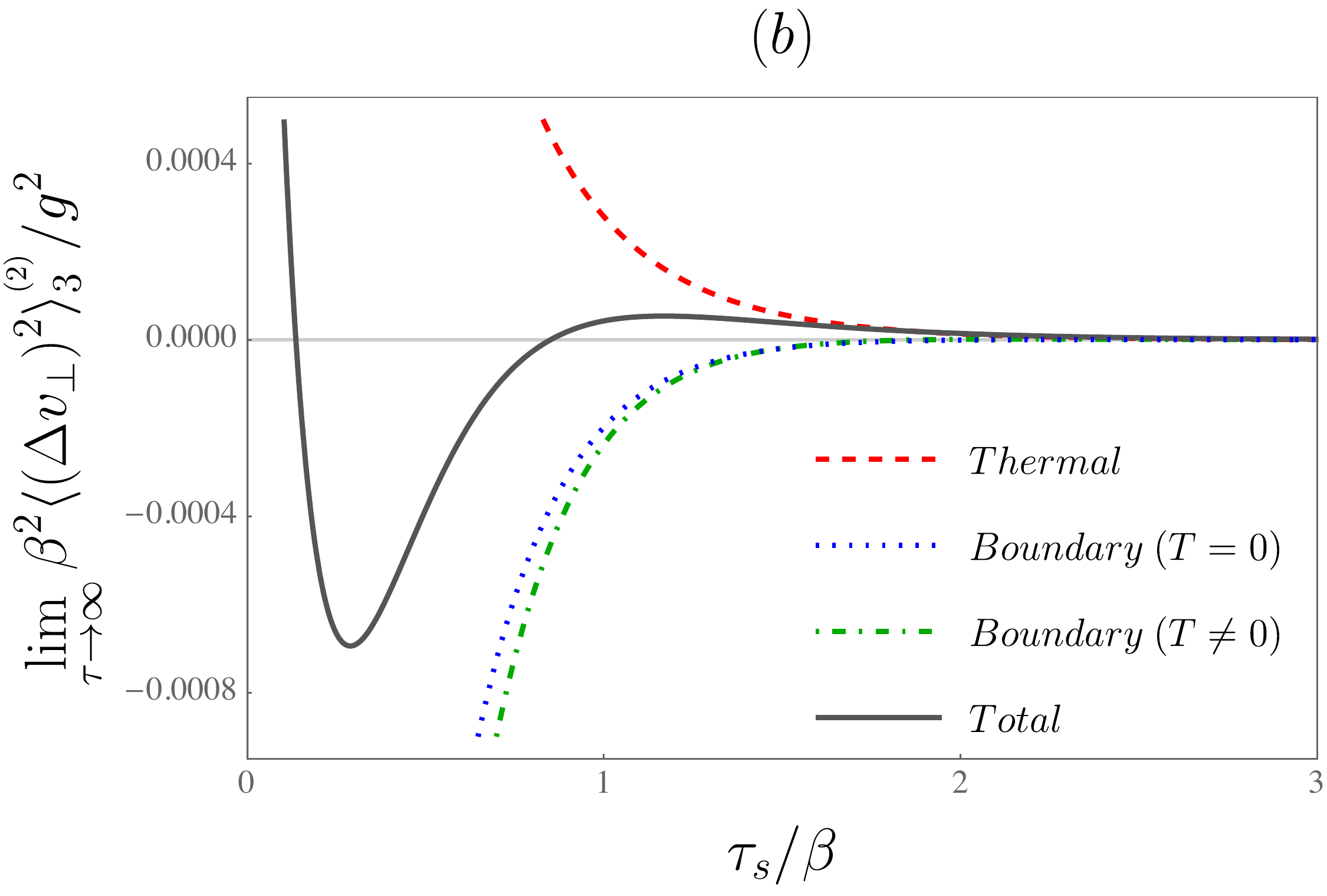}
\caption{Late-time behavior of the velocity dispersions for $D=3$ as function of the switching time. Here we set $mx=1$, and $\beta/x=1$. (a) Late-time dispersions for the parallel directions. (b) Late-time dispersions for the perpendicular direction. The boundary contribution is negative, opposing the thermal part, so that, for certain values of $\tau_s/\beta$ subvacuum effects appear.}
\label{fig7}
\end{figure}  
The late-time behavior of the dispersions as function of the switching time for $D=3$ are depicted in Fig.~\ref{fig7}, where the results in parallel and perpendicular directions are presented together with their partial contributions from thermal and boundary terms. Inspection of these figures show that the boundary contribution to the total dispersion is always a negative quantity, in opposition to the thermal contribution. That the dispersion related to presence of reflecting boundary can be negative is already known. However, the fact that temperature enhances the subvacuum effect related to the presence of the boundary is a quite interesting feature. 
This aspect can be clearly understood by comparing the magnitude of the dispersions due to the modified vacuum at zero temperature, depicted by the dotted curves in Fig.~(\ref{fig7}), with the modified vacuum at finite temperature, depicted by the dashed curves. As one can see, when the thermal bath is present the magnitude of the subvacuum effect related to the boundary contribution increases. 
The competition between these terms and the always positive thermal contribution, depicted by the dot-dashed curves in Fig.~(\ref{fig7}), produces a total dispersion (solid curves)  that can be positive or negative, depending on the specific value of $\tau_s$. The presence of the boundary plays a fundamental role in activating subvacuum effects -- no boundary no subvacuum behavior. However, as the boundary is implemented, the switching time will control the signal of the dispersions. For a short enough transition (with the limiting value $\tau_s = 0$ describing a sudden process) the dispersion will be positive and the particle will gain an amount of kinetic energy from the modified vacuum. However, for a certain range of values of $\tau_s$, subvacuum effects will occur, and the particle will lose part of its energy to the system. This is an effect that cannot be found in the realm of classical physics and is closely related to the renormalization process. 

Another aspect that can be inferred from the results in the \ref{appendix} is that the modified vacuum contribution dominates over the purely thermal contribution for higher field masses $m$. On the other hand, lowering $m$ or $\beta$ (raising the temperature) raises the dispersions until the thermal contribution will eventually dominate and no subvacuum effect will survive. Moreover, dispersions vanish when $\tau_s \to \infty$, as anticipated in the context of a transition described by the former sample function.

\section{Distance behavior of the velocity dispersions}
\label{secVI}
The introduction of a perfectly reflecting boundary at $x_1=0$ changes the topology of the space in which the field is defined by demanding that the field vanishes on the wall. Hence, when an expectation value is renormalized, by subtracting the divergent Minkowski contribution, a divergence appears at the wall position, in addition to the one for $\tau=2x$. Both divergences are regularized by implementing a smooth switching, as it suppresses the high-energy modes.

To investigate how the dispersions behave with the distance to the plate, note that, when $x/\beta\ll 1$, we have $(1-\gamma^2_l)^{1/2}\simeq (-\gamma^2_l)^{1/2}$. Hence the mixed contributions to the dispersions can be approximated as 
\begin{eqnarray}
\langle(\Delta v_{\parallel})^2\rangle^{(1)}_{D,\textrm{mixed}}&\simeq&-\langle(\Delta v_{i})^2\rangle^{(1)}_{D,\textrm{thermal}}
\nonumber \\
\langle(\Delta v_{\perp})^2\rangle^{(1)}_{D,\textrm{mixed}}&\simeq&\langle(\Delta v_{i})^2\rangle^{(1)}_{D,\textrm{thermal}}
-8\pi x^2\langle(\Delta v_{i})^2\rangle^{(1)}_{D+2,\textrm{thermal}}\,.
\nonumber
\end{eqnarray}

Thus, for the dispersion in the parallel direction, mixed and thermal contributions cancel near the wall, and only the temperature-independent modified vacuum term remains.
On the other hand, in the perpendicular direction mixed contribution equals the pure thermal one as $x\rightarrow 0$, and the dispersion grows with the temperature at the wall
A similar behavior is found for the massless vector field \cite{delorenci2019b}. 

Now, let us investigate vacuum versus thermal dominance. Let us define \cite{delorenci2019b} 
\begin{equation}
    \eta_{_{D}}=\left|\frac{\langle v^2 \rangle_{D,\beta}-\langle v^2 \rangle_{D,\infty}}{\langle v^2 \rangle_{D,\infty}}\right|
    \label{eq37}
\end{equation}
where $\langle v^2 \rangle_{D,\beta}$ is the mean squared velocity,
\begin{equation}
    \langle v^2 \rangle_{D,\beta}=\sum_i \langle v_i^2 \rangle_{D,\beta}=(D-1)\langle (\Delta v)^2_{\parallel}\rangle_{D,\beta}+\langle (\Delta v)^2_{\perp}\rangle_{D,\beta}.
\nonumber
\end{equation}
Then, when $\eta_{_{D}}>1$ thermal effects dominate, and if $\eta_{_{D}}<1$, vacuum effects dominate.
%

%

By rewriting Eq.~\eqref{eq37} for the late-time regime, using $F^{(2)}_{\tau_s,\tau}(t)$, and taking $x\rightarrow 0 $ we find
\begin{align}
  \lim_{\stackrel{\tau\rightarrow\infty}{x\rightarrow0}}\eta^{(2)}_{_{D}} 
	=\left|\frac{2\lim_{\tau\rightarrow\infty}\langle (\Delta v_i)^2\rangle^{(2)}_{D,\textrm{thermal}}}{(D-2) \lim_{\stackrel{\tau\rightarrow\infty}{x\rightarrow0}}\langle (\Delta v_{\parallel})^2\rangle^{(2)}_{D,\textrm{vacuum}}}\right|.
    \label{eq38}
\end{align}

Note that, for $D=2$, $\eta$ diverges, which shows the thermal dispersions always dominates near the wall in this case, independently on the magnitude of the field mass.


%

%
\begin{figure}[ht!]
\includegraphics[width=0.45\textwidth]{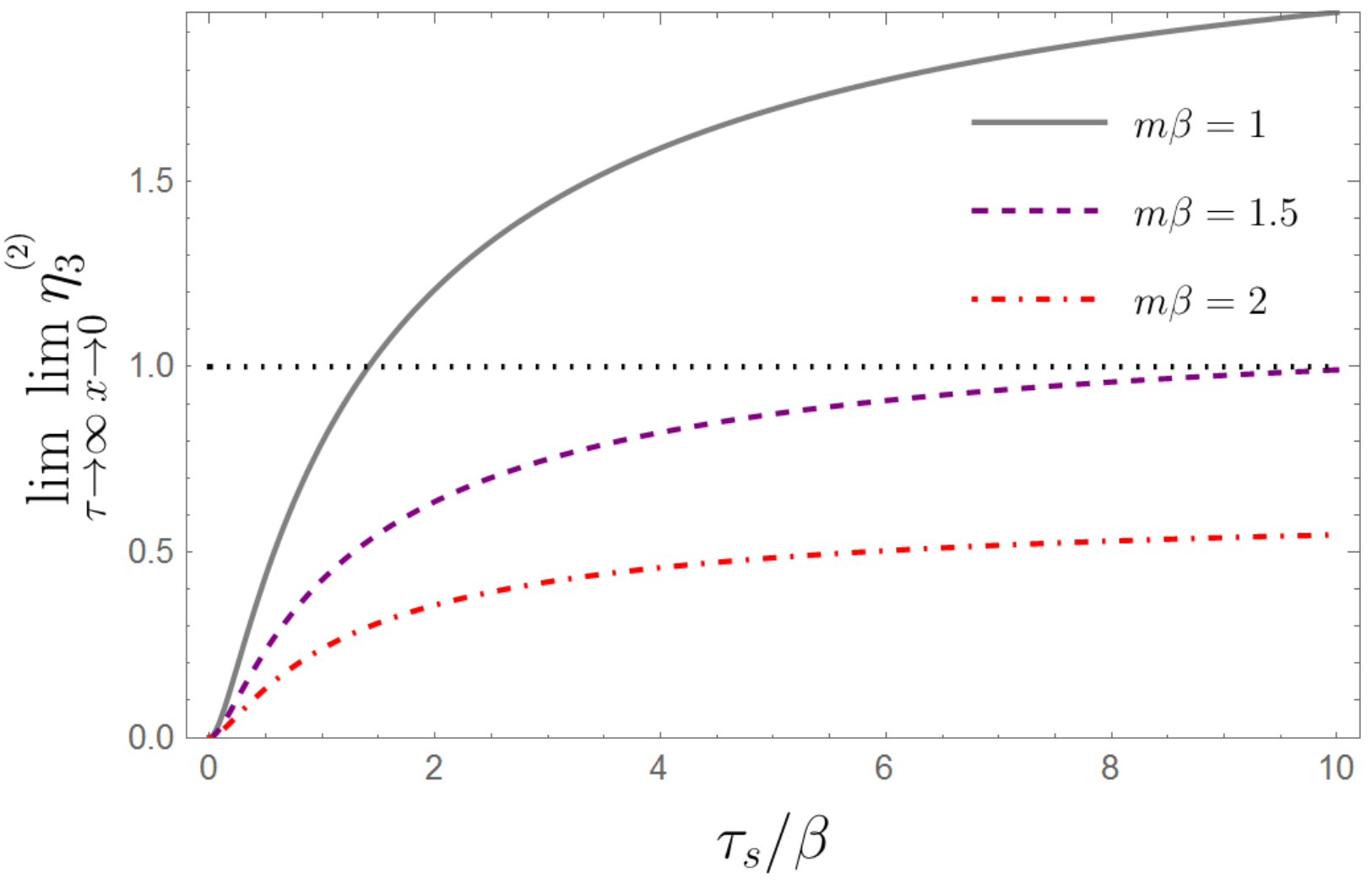}
\caption[Vacuum versus Thermal dominance in the wall for different values of $\tau_s/\beta$ and $m\beta$, with $D=3$.]{Vacuum versus thermal dominance near the wall for different values of $\tau_s/\beta$ and $m\beta$, with $D=3$. For high values of the field mass the vacuum contribution  always dominate, independently of the switching time.}
\label{fig10}
\end{figure}
This expression  vanishes in the limit of a sudden transition ($\tau_s/\beta\rightarrow0$), as there we have the divergence on the wall for the vacuum contribution. As $\tau_s/\beta$ increases, the divergence is regularized and the vacuum contribution to the dispersion is smoothed. Thus, in the limit of $\beta$ and $m$ going to zero $\eta_3$ grows as well, before it stabilizes for greater values of $\tau_s$. Notwithstanding, the field mass plays an important role, as depicted in Fig.~\ref{fig10}. It lowers the curve, and for masses higher than around $m\beta\simeq1.5$ the vacuum term dominates for any switching time. Such behavior comes from the suppression of the thermal contribution due to the mass of the field. 
\begin{figure}[h!]
    \includegraphics[width=0.45\textwidth]{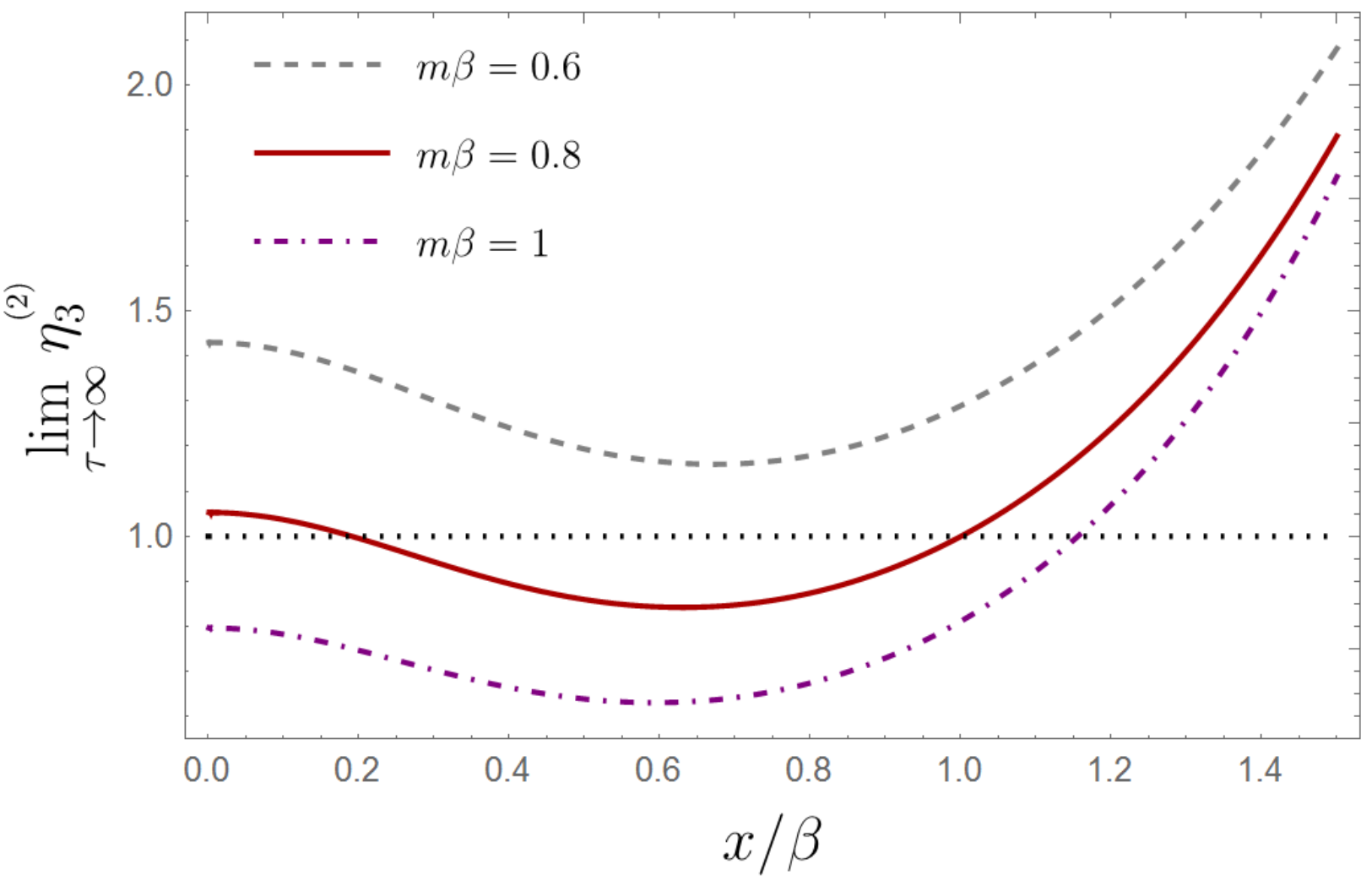}
\caption{Distance behavior of the vacuum versus thermal dominance for different values of $m\beta$, here $D=3$ and $\tau_s/\beta=1$. When $m\beta=0.6$ thermal effects always dominate. Then, as mass is increased, vacuum effects dominate in certain regions.}
\label{fig11}
\end{figure}
\begin{figure}[h!]
    \includegraphics[width=0.45\textwidth]{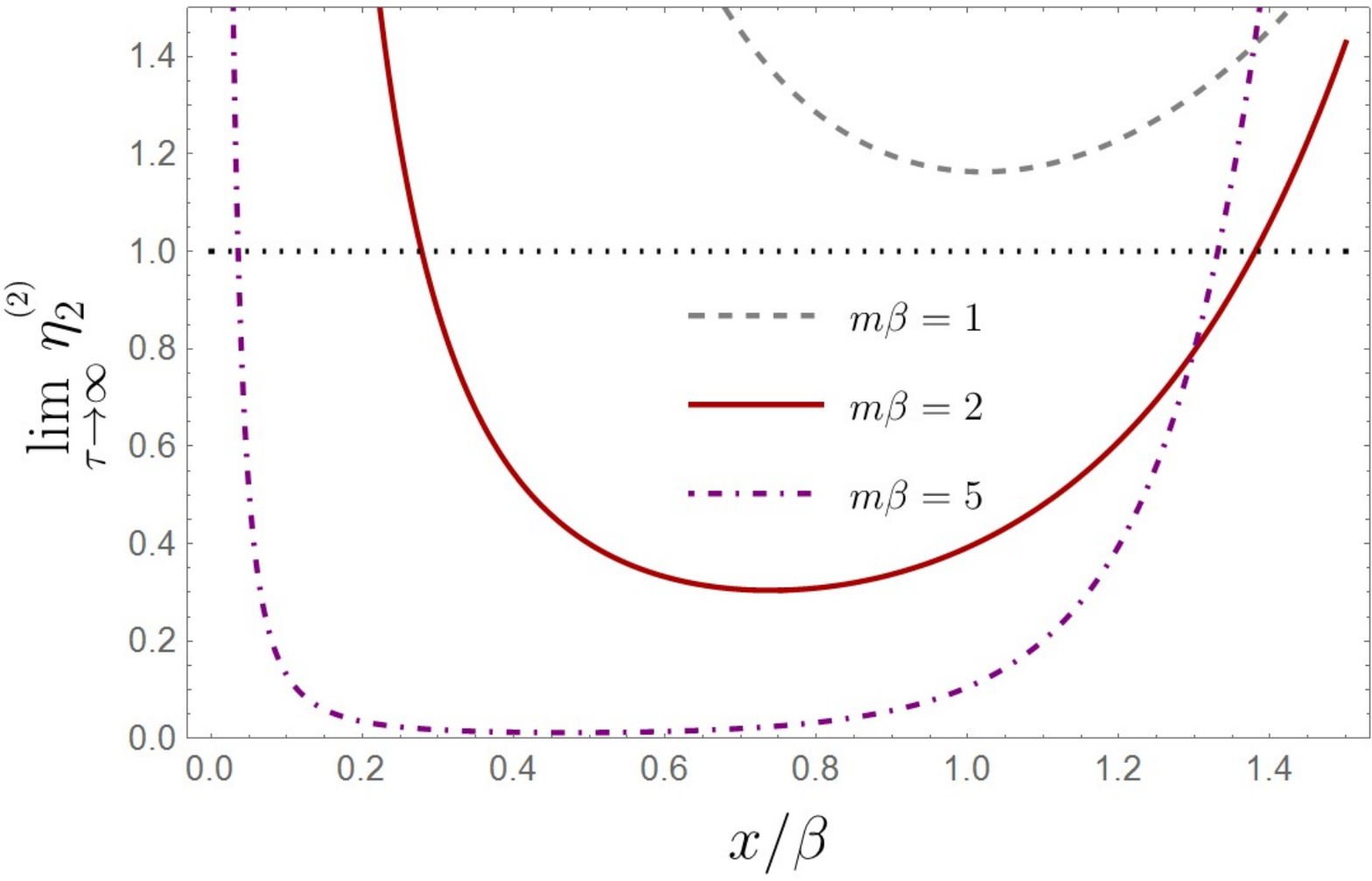}
\caption[Distance behavior of the vacuum versus thermal dominance for different values of $m\beta$, here $D=2$ and $\tau_s/\beta=1$.]{Distance behavior of the vacuum versus thermal dominance for different values of $m\beta$. Here we set $D=2$ and $\tau_s/\beta=1$. In this case we always have thermal dominance sufficiently near the wall. However, vacuum dominance occurs in certain regions, depending on the value of the field mass. }
\label{fig12}
\end{figure}

Further, we investigate the distance behavior of $\eta$ as $\tau\rightarrow\infty$. As expected, deep in the bulk  ($x/\beta\rightarrow\infty$) we found that $\eta\rightarrow\infty$, i.e., the boundary effects no more contribute to the dispersions. When $D=3$, from Fig.~\ref{fig11} we see that the dominance is dependent on the mass, for $m\beta=0.6$ only thermal effects dominate, then, for $m\beta=1$ the vacuum dominates near the wall. However, for intermediary values of the mass, as $m\beta=0.8$, the behavior oscillates: first the thermal part, then the vacuum, and finally the thermal contribution dominates.  

For $D=2$ thermal effects dominate near the wall for any value of the mass, as shown in  Fig.~\ref{fig12}. As distance to the wall gets larger, vacuum contribution dominates for high values of $m$, after which thermal effects dominate deep in the bulk. As before, above a certain value of $m\beta$ thermal effects dominate at any distance to the wall.

Finally, it is worth to mention that when $D\neq2$  and  $m\rightarrow0$, Eq. (\ref{eq38})  reduces to half the result found for the electromagnetic case \cite{delorenci2019b}, as expected.

\section{Final remarks}
The stochastic motion induced by quantum fluctuations of a massive scalar field at finite temperature was here examined. The non-Huygensian feature of a massive field creates an oscillatory pattern in the dispersions of the particle velocity, in agreement with previous investigations at zero temperature  \cite{Camargo2019}. In the presence of a thermal bath, effects linked to the field mass oppose those associated with temperature -- at higher temperatures the oscillatory pattern is suppressed, and for larger field mass the thermal dispersions get lower. Such interplay between mass and temperature becomes even more relevant near the reflective boundary. For fields with higher values of mass, the distance to the plate for which the pure vacuum contribution to the dispersions dominates over thermal ones increases, and in some cases an oscillation can be seen in this dominance, as the wall is approached. Here it is demonstrated that in the presence of a boundary, subvacuum effects leads to negative values of the dispersions even at finite temperatures. Particularly, it is shown that temperature enhances the contribution to the subvacuum effects related to the presence of the boundary, which is a remarkable feature as the pure thermal part can be detached from the boundary contribution. Hence, raising the temperature improves the measurements of negative dispersions.

\begin{figure}[ht!]
\includegraphics[width=0.45\textwidth]{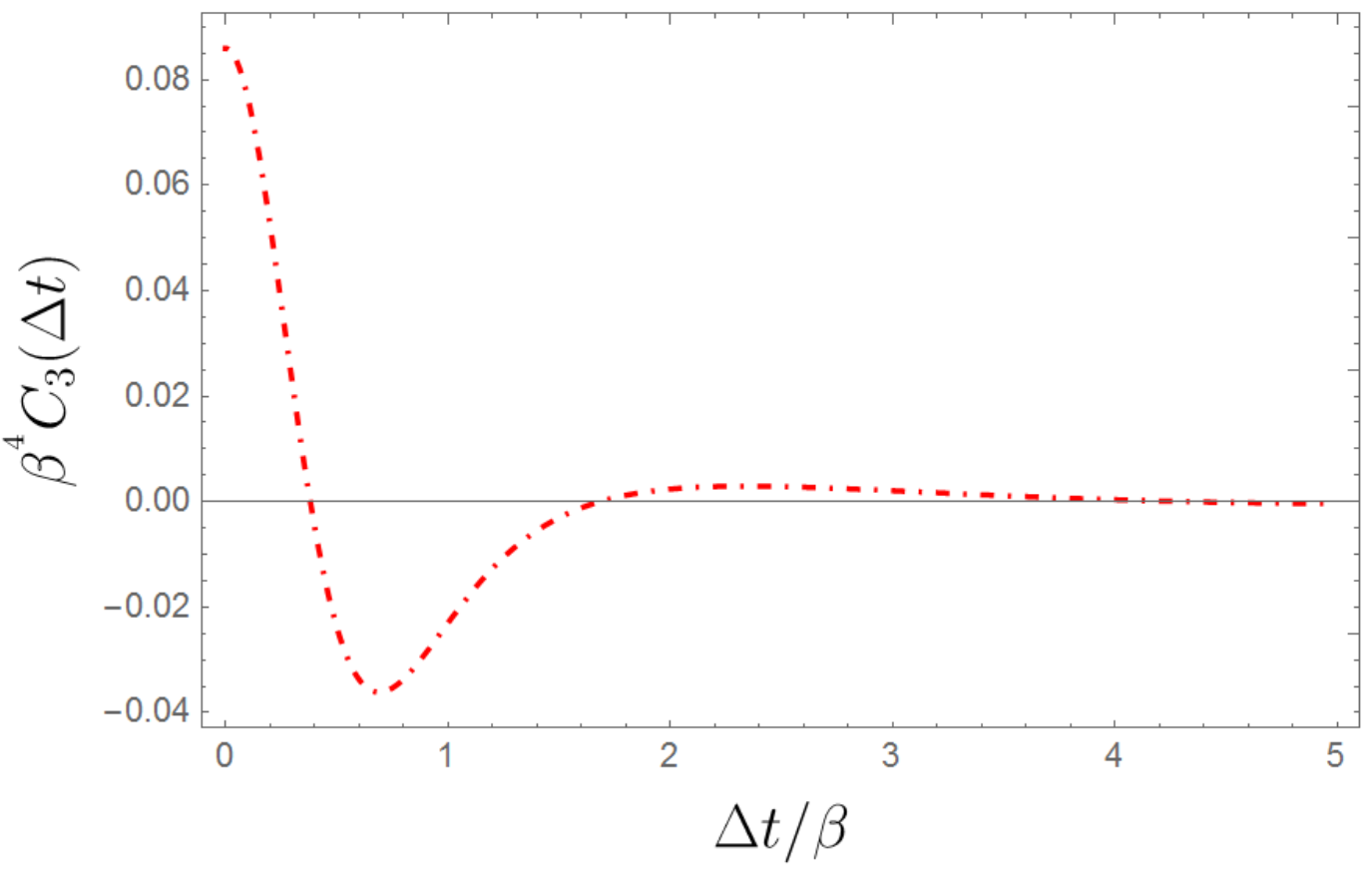}
\caption[Correlation function for the thermal contribution to the velocities dispersions, here we have $m\beta=1$ and $D=3$]{Correlation function for the thermal contribution to the velocities dispersions for $D=3$. Here we set  $m\beta=1$. One can see that the correlation takes on negative values, so that the dispersions become bounded for large values of the interaction time.}
\label{fig14}
\end{figure}

The nature of the stochastic motion induced by field fluctuations is such that, even in the absence of a dissipative force, the dispersions are bounded. Moreover, we have seen that for late-times the dispersions depend only on the switching time, so the quadratic velocity dispersions do not go to its equipartition value even in the pure thermal case. Henceforth, the time motion here studied is different from the usual Brownian motion. This difference comes from the correlation function $C(\Delta t)$, through which the dispersions are calculated as
\begin{equation*}
    \langle(\Delta v_i)^2\rangle=g^2\int_{0}^{\tau}\mathrm{d} t\int_{0}^{\tau}\mathrm{d}t'C(\Delta t).
\end{equation*}
For the usual Brownian motion the correlation function $C(\Delta t)$ is non negative and decays monotonically with $\Delta t$ with the relaxation time of the fluid, giving  $\langle(\Delta v_i)^2\rangle\propto\tau$, the usual random walk motion, and a dissipative force is needed \cite{ford2005}. For the boundary contributions this behavior is expected, as then the correlation function is known to have negative values. However, we saw that it occurs even for the pure thermal contributions. In such case the correlation function is
\begin{multline}
    C_{_{D}}(\Delta t)=\frac{2}{\beta^{D+1}}\textrm{Re}\,\sum_{l=1}^{\infty}\left[\frac{m\beta}{2\pi \sqrt{-(\Delta t/\beta+il)^{2}}}\right]^{\frac{D+1}{2}}\\\times K_{_{\frac{D+1}{2}}}\left[m\beta\sqrt{-(\Delta t/\beta+il)^{2}}\right],
    \nonumber
\end{multline}
which, as depicted in Fig.~\ref{fig14}, also takes on negative values, making the dispersions bounded for $\tau\rightarrow\infty$, just as for the boundary contribution. 

Regarding the assumptions made in our model, for the particle position dispersions to be neglected we must have $\braket{(\Delta x_i)^2}=\int_0^{\tau}\mathrm{d} t\int_0^{\tau} \mathrm{d} t'\braket{v_i(t)v_i(t')}\ll x^2_i$. The regime in which this condition is satisfied was explored for fluctuations of the electromagnetic field at zero temperature \cite{ford2004,delorenci2016} and at finite temperature \cite{hongwei2006}. Nonetheless, the velocity dispersions for the massless case acts as an envelope for the massive field. Henceforth, as the position dispersions is an integral of velocity correlations, it is expected that it has a greater value for the massless field. Then, the validity regime for massless fields should include the massive case. 
Another remark is that such an assumption may imposes constrains on lengthy interaction times. However, as we have seen, after an initial peak the dispersions oscillate around their late-time value, rapidly approaching it. Thus, there can be values of the time which satisfy the assumptions and in which the system is already in its late-time regime.

Further, dispersions were calculated by using Dirichlet's boundary condition. The results can be drastically changed if Neumann boundary conditions are used. It is was shown in \cite{delorenci2015} that in such case the mixed and vacuum terms change by an overall sign. Hence, we could not have subvacuum effects for the parallel directions, only in the perpendicular one, as there is not only a valley, but also a peak in the boundary contributions.

Concluding, some words about the negative dispersions are in order. The dispersion of a quantity measures how much it deviates from its mean value. Hence, it taking a negative value looks counterintuitive. What happens is that the dispersion of the particle velocities are lessen when compared to their value when the interaction starts, opposing the usual behavior of a free particle wave packet spreading \cite{ford2004}. Moreover, classical contributions from the interaction of the particle with the boundary must be taken in account in a more complete scenario.

\begin{acknowledgements}
This work was partially supported by the Brazilian research agencies CNPq (Conselho Nacional de Desenvolvimento Cient\'{\i}fico e Tecnol\'ogico) under Grant No. 305272/2019-5, and CAPES (Coordena\c{c}\~ao de Aperfei\c{c}oamento de Pessoal de N\'{\i}vel Superior).
\end{acknowledgements}

\appendix
\section{Calculations for the late-time regime}
\label{appendix}

Substituting the Fourier transform of $F^{(2)}_{\tau_s,\tau}(t)$ in Eq.~\eqref{eq13} gives for the thermal contribution
\begin{align}
    \langle(\Delta v_i)^2\rangle^{(2)}_{D,\textrm{thermal}}=&g^2\lim_{\x\rightarrow \x'}\Bigg[\frac{\partial}{\partial x_i}\frac{\partial}{\partial x'_i}\frac{2}{(2\pi)^{\frac{D}{2}}|\Delta \x|^{\frac{D}{2}-1}}
\nonumber\\&\times
		\sum_{l=1}^{\infty}\int_{0}^{\infty}\mathrm{d} k\frac{k^{\frac{D}{2}}}{\omega^3}[1-\textrm{cos}(\omega\tau)]\nonumber\\&\hspace{8mm}\times\textrm{e}^{-\omega(2\tau_s+l\beta)}J_{\frac{D}{2}-1}(k|\Delta \x|)\Bigg].
    \label{eq24}
\end{align}
By taking the limit $\tau\rightarrow\infty$ the integration over the term multiplied by $\textrm{cos}(\omega\tau)$ vanishes \cite{Camargo2019}. The remaining integral can presented as,
\begin{align}
    &\frac{2}{(2\pi)^{\frac{D}{2}}|\Delta {\bf x}|^{\frac{D}{2}-1}}\int_{0}^{\infty} \mathrm{d} k\frac{k^{\frac{D}{2}}}{\omega^3}\textrm{e}^{-\omega(2\tau_s+l\beta)}J_{\frac{D}{2}-1}(k|\Delta {\bf x} |)\nonumber\\&
   =-\frac{2(2\tau_s+l\beta)}{(2\pi)^{\frac{D}{2}}| \Delta {\bf x} |^{\frac{D}{2}-1}}\int_{0}^{\infty}\mathrm{d} k\frac{k^{\frac{D}{2}}}{\omega^2}\textrm{e}^{-\omega(2\tau_s+l\beta)}J_{\frac{D}{2}-1}(k|\Delta \bf{x}|)\nonumber\\&
  -\frac{1}{m}\frac{\partial}{\partial m}\left[\frac{2}{(2\pi)^{\frac{D}{2}}|\Delta {\bf x}|^{\frac{D}{2}-1}}\int_{0}^{\infty} \mathrm{d} k\frac{k^{\frac{D}{2}}}{\omega}\textrm{e}^{-\omega(2\tau_s+l\beta)}J_{\frac{D}{2}-1}(k|\Delta \bf{x}|)\right].
\nonumber
\end{align}
The second integral on the right hand side of this expression is already known \cite{gradshteyn}, and the first one, after taking the spatial derivatives and the limit $\x'\rightarrow\x$ is identified with generalized Hypergeometric functions ${}_1F_2$. Hence, the late-time behavior of the thermal dispersion is
\begin{align}
\lim_{\tau\rightarrow\infty}\langle(\Delta v_i)^2\rangle^{(2)}&_{D,\textrm{thermal}}=
\nonumber \\&\frac{2g^2}{\pi x^{D-1}}\sum_{l=1}^{\infty}\left[\left(\frac{mx}{4\pi\alpha_l}\right)^{^{\frac{D-1}{2}}}K_{_{\frac{D-1}{2}}}\left(2mx\alpha_l\right)\right.
\nonumber\\
&\hspace{5mm}-\left.\frac{\alpha_l(mx)^{D}}{2^{D-1}\pi^{\frac{D}{2}-1}\Gamma(\frac{D}{2}+1)}I(D,mx\alpha_l)\right].
\label{eq25}
\end{align}
Here we have defined $\alpha_l \doteq \tau_s/x+l\beta/2x$, and
\begin{align}
    &I(D,y)\doteq 
		-\frac{\pi}{2}\textrm{cosec}\left(\frac{\pi}{2}D\right)-\frac{y}{2\sqrt{\pi}}\Gamma\left(-\frac{D}{2}-1\right)\nonumber\\
    &\times\Gamma\left(\frac{D}{2}+1\right){}_1{F}_2\left[1/2;3/2,(D+3)/2;y^2/4\right]\nonumber\\
		&+\frac{1}{y^D}\Gamma(D){}_1F_2\left[-D/2;(1-D)/2,1-D/2;y^2/4\right].
\nonumber
\end{align}
The distante to the wall $x$ appearing in Eq.~(\ref{eq25}) is fictitious. It was introduced only to make easy its comparison with the other dispersions, as for instance in the figures of Secs.~\ref{secV} and \ref{secVI}. The thermal dispersion is clearly isotropic.  

The massless case with $D=3$ leads to  the expression found for the electromagnetic case in Ref. \cite{delorenci2019b}, up to a 1/2 factor. 
 And for $D=2$
\begin{equation}
     \lim_{\tau\rightarrow\infty}\langle(\Delta v_i)^2\rangle^{(2)}_{2,\textrm{thermal}}\stackrel{m\rightarrow0}{=}\frac{g^2}{2\pi\beta}\sum_{l=1}^{\infty}\frac{1}{[l+2(\tau_s/\beta)]},
\nonumber
\end{equation}
which is a divergent summation, which announces the well known fact that no thermal equilibrium is possible in this case. At this point we should notice that if we have insisted in using the results obtained by implementing $F^{(1)}_{n,\tau}(t)$ to study the late-time behavior of the dispersions, we would obtain an apparent regularization of the infrared divergence when $D=2$, as one can directly inspect in Eq. (\ref{eq16}). However such aspect would be just a consequence of setting a long-lasting transition time. In fact, there is no transition when $\tau_s\to\infty$.

The mixed contribution is obtained in the same way as done in obtaining Eq.~\eqref{eq24}, but exchanging $\Delta\x\rightarrow\hat{\Delta}\x$. Now, the argument of the Bessel function will not vanish in the limit of point coincidence. Thus, after taking the spatial derivatives and $\x'\rightarrow\x$ we obtain
\begin{align}
    &\lim_{\tau\rightarrow\infty}\langle(\Delta v_\parallel)^2\rangle^{(2)}_{D,\textrm{mixed}}=\nonumber\\
		&\frac{-2g^2}{\pi x^{D-1}}\sum_{l=1}^{\infty}\left\{\left[\frac{mx}{4\pi\sqrt{1+\alpha^2_l}}\right]^{^{\frac{D-1}{2}}}K_{_{\frac{D-1}{2}}}\left(2mx\sqrt{1+\alpha^2_l}\right)\right.
    \nonumber\\
    & -\frac{\alpha_l (mx)^{\frac{D}{2}}}{2^{D-1}\pi^{\frac{D}{2}-1}}\nonumber\\
		&\times\left. \int_{1}^{\infty} \mathrm{d} u \frac{(\sqrt{u^2-1})^{\frac{D}{2}}}{u}\textrm{e}^{-2mx\,\alpha_l}J_{\frac{D}{2}}\left(2mx\sqrt{u^2-1}\right)\right\},\nonumber 
     \end{align}
    \begin{align}    
    \lim_{\tau\rightarrow\infty}\langle(\Delta v_\perp)^2\rangle^{(2)}_{D,\textrm{mixed}}=&8\pi x^2\lim_{\tau\rightarrow\infty}\langle(\Delta v_\parallel)^2\rangle^{(2)}_{D+2,\textrm{mixed}}\nonumber\\
		&-\lim_{\tau\rightarrow\infty}\langle(\Delta v_\parallel)^2\rangle^{(2)}_{D,\textrm{mixed}}.\hfill
\nonumber
\end{align}

The vacuum contribution is again just half the mixed part with $l=0$,
\begin{align}
    &\lim_{\tau\rightarrow\infty}\langle(\Delta v_\parallel)^2\rangle^{(2)}_{D,\textrm{vacuum}}=-\frac{g^2}{\pi x^{D-1}}\nonumber\\
		&\times\Bigg\{\left[\frac{mx}{4\pi\sqrt{1+(\tau_s/x)^2}}\right]^{^{\frac{D-1}{2}}}K_{_{\frac{D-1}{2}}}\left(2mx\sqrt{1+(\tau_s/x)^2}\right)\nonumber\\
    &-\frac{(\tau_s/x) (mx)^{\frac{D}{2}}}{2^{D-1}\pi^{\frac{D}{2}-1}}\nonumber\\
		&\times\int_{1}^{\infty} \mathrm{d} u\frac{(\sqrt{u^2-1})^{\frac{D}{2}}}{u}\textrm{e}^{-2mx(\tau_s/x)}J_{\frac{D}{2}}\left(2mx\sqrt{u^2-1}\right)\Bigg\},
\nonumber
    \end{align}
    \begin{align}
    \lim_{\tau\rightarrow\infty}\langle(\Delta v_\perp)^2\rangle^{(2)}_{D,\textrm{vacuum}}=&8\pi x^2\lim_{\tau\rightarrow\infty}\langle(\Delta v_\parallel)^2\rangle^{(2)}_{D+2,\textrm{vacuum}}\nonumber\\
		&-\lim_{\tau\rightarrow\infty}\langle(\Delta v_\parallel)^2\rangle^{(2)}_{D,\textrm{vacuum}}.\hfill
\nonumber
\end{align}
%



\end{document}